\documentclass[]{article}
\usepackage{graphicx} 
\usepackage{amsmath}
\usepackage{amssymb,commath}
\usepackage{tikz}
\usepackage{graphicx}
\usepackage[footnotesize,hang]{caption}
\usepackage{subfig}
\usepackage{color}
\usepackage{rotating}
\usepackage{mathtools}
\usepackage{xcolor}
\usepackage{epstopdf}
\usepackage{cite}
\usepackage{url}
\usepackage{pifont}
\usepackage[margin=0.85in]{geometry}

\title{Recursive behavior in a diatomic FPUT lattice}
\author{Guo Deng$^1$, Andrea Pezzi$^{2}$, Genghong Lin$^{1,3}$,  Miguel Onorato$^{2,4}$}
\date{%
    $^1$School of Mathematics and Information Science, Guangzhou University, Guangzhou 51006, China\\[2ex]%
    $^2$Dipartimento di Fisica, Università degli Studi di Torino, Via Pietro Giuria 1, 10125 Torino, Italy\\[2ex]%
    $^3$Guangzhou Center for Applied Mathematics, Guangzhou University, Guangzhou, 510006, China\\[2ex]%
    $^4$INFN, Sezione di Torino, Via Pietro Giuria 1, 10125 Torino, Italy\\[2ex]%
}

\begin{document}

\maketitle

\begin{abstract}
We study the diatomic Fermi–Pasta–Ulam–Tsingou (FPUT) lattice with cubic anharmonic potential, and analyze the recurrent behaviour of its solutions. We find that two distinct types of recurrence occur. One type is the classic FPUT recurrence;  for such recurrence, we find that the relation between recurrence period and nonlinear strength is similar to that in the monatomic case. The other type, which cannot exist in the monatomic lattice, is the recurrence due to the interactions between modes in the two branches of the dispersion relation. Indeed, we prove the existence of the optical-acoustical-acoustical resonant interaction between three Fourier modes for which a recurrent behavior in the distribution of the energy is observed.  In addition, we develop a reduced Fourier-space dynamical model that reproduces the same recurrent behavior. 
We assess the robustness of our results through numerical simulations of the diatomic Toda lattice and the diatomic granular chain; in both cases, the same recursive behavior is observed.
Finally, in the continuous limit, we derive from the diatomic model a system of three coupled PDEs which are known to be integrable. 
\end{abstract}

\section{Introduction}
To investigate the phenomenon of thermalization, Fermi and his collaborators \cite{fermi1955} numerically studied a monatomic chain with potential consisting of a harmonic term plus a cubic or a quartic contribution, the $\alpha-$ and $\beta-$ FPUT lattice, respectively~\cite{fermi1955}. 
The motivation to include a nonlinear term is that in linear systems there is no intrinsic mechanism leading to energy sharing among Fourier modes and thermalization of the normal modes is never possible. Fermi and collaborators initially excited only the fundamental Fourier mode, and they expected that, because of the presence of  nonlinearity, energy would evenly distribute  among all the Fourier modes. However, surprisingly, instead of thermalization they observed the recurrence of energy, which is typical for integrable systems. Their pioneering numerical simulations have brought numerous insights to various research avenues, including integrable systems~\cite{ZabuskyKruskal1965,TodaBook}, solitary wave dynamics~\cite{FrieseckeWattis,FrieseckePego}, thermal properties of nonlinear lattices~\cite{ZhangZhao,lepri2016thermal} and energy recurrence~\cite{Campbell2019,Porter2023}. 

With the advancement of computational power, it has become possible to simulate the long-time dynamics of the FPUT lattice. These studies have shown that the recurrence first observed in \cite{fermi1955} does not persist indefinitely, as the system eventually approaches thermal equilibrium~\cite{Benettin2013}. Within the framework of wave turbulence theory~\cite{nazarenko2011,zakharov2012}, it has been given evidence that the interactions among Fourier modes satisfying the resonance conditions enable effective energy exchange, ultimately driving the thermalization of nonlinear systems~\cite{onorato2015,lvov2018,pistone2018}. Evidence has also shown that, in the large-box limit, the mechanism responsible for the thermalization of both $\alpha$- and $\beta$-FPUT lattices is the four-wave resonant interaction process~\cite{onorato2015,lvov2018}.

A substantial body of research on the energy sharing and thermal properties of nonlinear lattices has focused on unperturbed monatomic systems. In contrast, diatomic lattices exhibit several  properties that differ significantly from those of monatomic lattices. 
The difference in thermal properties between monatomic and diatomic lattices is largely due to the difference in their dispersion relation~\cite{Fu2019}. As a result, the diatomic chain can display exact three-wave resonant interactions~\cite{pezzi2025multi}, which are not possible in the monatomic case. In addition, optical Fourier modes~\cite{Maiocchi2019,Galgani1992,Bambusi1993} can remain frozen over certain ranges of the system’s mass ratio. 

In this work, we investigate the recurrence properties of the diatomic $\alpha$-FPUT chain. As in the original study by Fermi and collaborators, we consider an initial excitation consisting of a long-wavelength mode belonging to the acoustic branch and observe recurrent dynamics analogous to those found in the classical $\alpha$-FPUT problem.
However, this is not the only type of recurrence exhibited by the system. Owing to resonant interactions between modes in the acoustic and optical branches, we find that an initial condition with energy concentrated in a single optical mode can transfer energy to acoustic modes and, after a sufficiently long time, recover its initial energy. Remarkably, this recurrence persists over very long timescales, indicating a robust exchange mechanism between the two branches. 
This paper is organized as follows. In Section~\ref{s:model}, we introduce the diatomic $\alpha$-FPUT model. In Section~\ref{s:fput recurrence}, we numerically show that the FPUT recurrence persists in the diatomic model when the fundamental Fourier mode is initially excited. In Sections~\ref{s:recurrence due to resonance}, we demonstrate the existence of resonant interactions among three Fourier modes and show that these interactions give rise to a different type of energy recurrence. In Section~\ref{sss:Fourier}, a reduced system of ODEs describing the dynamics of the three modes is derived, and its solutions are compared with those of the full system. In Section~\ref{s:linear limit}, we investigate the linear limit of the recurrence phenomenon in the diatomic lattice. In Section~\ref{s:variance}, we examine the robustness of the energy recurrence by studying the Toda and the granular lattices; fluctuations in the particle masses are also considered. The derivation of an integrable system of three coupled PDEs is presented in Section~\ref{Sec:threewave}. Finally, conclusions are presented in Section~\ref{s:conclusion}.

\section{The diatomic $\alpha-$ FPUT model}
\label{s:model}
\begin{figure}[htbp]
\begin{center}
\includegraphics[width=0.49\textwidth]{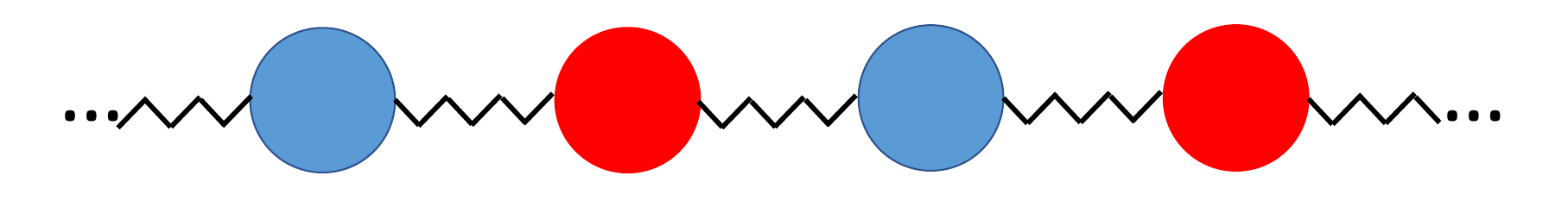}
\end{center}







    
    \caption{Lattice with alternating masses. Different colors represent particles with different masses.}
    \label{f:diatomic}
\end{figure}

We study a chain with alternating masses as illustrated in Figure~\ref{f:diatomic}. The chain has $2N$ particles, the light mass is denoted by $m_1$ and the heavy mass is indicated by $m_2$. 

For the $n-$th particle, the equilibrium position is $y_{0,n}=na$, where $a$ is the lattice spacing; the displacement with respect to its equilibrium is denoted by $y_n$, and the momentum is $p_n$. We impose periodic boundary condition, with which $y_0=y_{2N}$. Besides standard Hooke
forces between neighbouring masses, we include nonlinear forces, i.e., an anharmonic potential. The Hamiltonian
takes the following form:
\begin{equation}
	H =\frac{1}{2}\sum_{n=1}^{N}(m_1\dot y _{2 n}^2+m_2\dot y _{2 n-1}^2)
	+\frac{\kappa}{2} \sum_{n=0}^{2N-1}  (y_{n+1}-y_{n})^2+  \frac{\alpha}{3} \sum_{n=0}^{2N-1}  (y_{n+1}-y_{n})^3,
	 \label{cubicpot1} 
\end{equation}
where $\kappa$ and $\alpha$ quantify the linear and the nonlinear strength of the system, respectively. 
The equations of motion for particles with mass $m_1$ and $m_2$ are given as
\begin{equation}
m_1\ddot{y}_{2n}=\kappa(y_{2n+1}+y_{2n-1}-2y_{2n})
+\alpha \bigl[(y_{2n+1}-y_{2n})^2-(y_{2n}-y_{2n-1})^2 \bigr]
\label{e:motion1}
\end{equation}
and
 \begin{equation}
m_2\ddot{y}_{2n+1}=\kappa(y_{2n+2}+y_{2n}-2y_{2n+1})
+\alpha \bigl[(y_{2n+2}-y_{2n+1})^2-(y_{2n+1}-y_{2n})^2 \bigr].
\label{e:motion2}
\end{equation}

\begin{figure}[b!]
\centering
\centerline{
\includegraphics[width=0.49\textwidth]{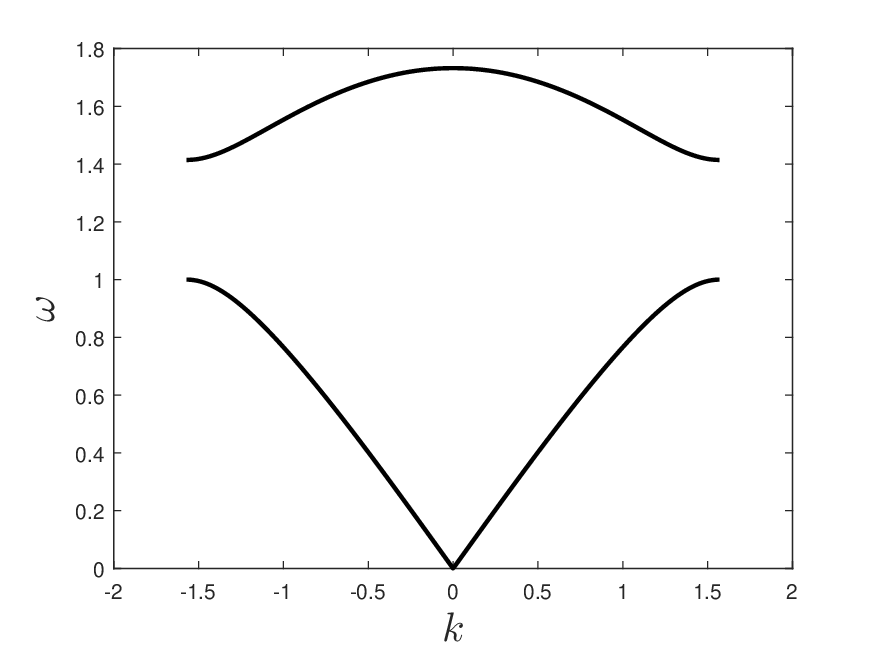}}





    \caption{Dispersion relation in linear diatomic lattice, where the linear coefficient $\kappa=1$, light mass $m_1=1$ and heavy mass $m_2=2$. The upper branch is the optical branch and the lower branch is the acoustical branch. }
    \label{f:dispersion}
\end{figure}
To find the dispersion relation, we set $\alpha=0$ and we search for solutions to~\eqref{e:motion1}-\eqref{e:motion2} of the form
\begin{equation}
\begin{split}
&y_{2n}(t) = 
	A_k e ^ {i(2naq_k-\omega_k t)} \\
&	y_{2n+1}(t) = B_k e ^{i((2n+1)aq_k-\omega_kt)},
\end{split}
\label{e:harmonic solution}
\end{equation}
where $\omega_k$ is the angular frequency, and $q_k$ are the discrete wave numbers:
\begin{equation}
\label{wavenumber}
\textcolor{black}{
q_k=\frac{\pi k}{Na} \qquad k \in \bigg(\!-\frac{N}{2},\frac{N}{2}\bigg] \cap \mathbb{Z}.
}
\end{equation}
Wee obtain two branches of the dispersion relation
\begin{equation}
\label{omega2}
\omega_{\pm,k} ^2= \kappa \frac{m_1+m_2}{m_1m_2} \Bigg[1\pm 
\sqrt{1-\frac{4m_1m_2}{(m_1+m_2)^2}\sin^2(aq_k)} \Bigg];
\end{equation}
$\omega_{\pm,k}$ are known as the optical and acoustical branches, respectively. In  Figure~\ref{f:dispersion}, we show the dispersion relation for the diatomic lattice.
The coefficients $A_k$ and $B_k$ in Equation~\eqref{e:harmonic solution} satisfy the following relation
\begin{equation}
\label{betarelation}
\frac{B_k}{A_k}=
\frac{2\kappa-m_1\omega^2_{\pm,k}}{2\kappa\cos(aq_k)}=
\frac{2\kappa\cos(aq_k)}{2\kappa-m_2\omega^2_{\pm,k}}\equiv\beta_{\pm,k}.
\end{equation}

Due to the coupling between the even--odd sub--lattices, in a diatomic chain a simple direct Fourier transform is not sufficient to diagonalize the linear operator. \textcolor{black}{
Using the identity $\beta_{k,+} \beta_{k,-}=-m_1/m_2$ and defining the effective modal masses
}
\begin{equation}
\textcolor{black}{
m^{\pm}_{12,k}=m_1+m_2\,\beta_{\pm,k}^2,
}
\label{e:m12}
\end{equation}
\textcolor{black}{
we introduce the following transformation, which projects the system onto the optical $(+)$ and acoustic $(-)$ modes
}
\begin{equation}
    \textcolor{black}{
    Q_{\pm,k}(t)=\frac{m_1}{m_{12,k}^\pm}\sum_{n=1}^N y_{2n}\mathrm{e}^{i2naq_k}+
    \frac{m_2 \beta_{\pm,k}}{m_{12,k}^\pm}\sum_{n=0}^{N-1} y_{2n+1}\mathrm{e}^{i(2n+1)aq_k}.
    }
    \label{e:ft}%
\end{equation}
The inverse transform of~\eqref{e:ft} is given by

\begin{subequations}
\begin{align}
    y_{2n}(t) &= 
    \textcolor{black}{
        \frac{1}{N} \sum_k \big( Q_{+,k}(t) + Q_{-,k}(t) \big)\, \mathrm{e}^{-i 2n a q_k}
    },\\
    y_{2n+1}(t) &= 
    \textcolor{black}{
        \frac{1}{N} \sum_k \big( \beta_{+,k} Q_{+,k}(t) + \beta_{-,k} Q_{-,k}(t) \big)\, \mathrm{e}^{-i (2n+1)a q_k}
    }
\end{align}
\label{e:transform}
\end{subequations}

The physical interpretation of \eqref{e:transform} is to write the displacement of each particle as a superposition of all possible lattice waves. 
Substituting~\eqref{e:transform} into~\eqref{cubicpot1}, we obtain the Hamiltonian whose quadratic part is in diagonalized form:
\begin{eqnarray}
NH&=&\sum_{k,s=\pm} \biggl\{\frac{\lvert {P}_{s,k} \rvert^2}{2 m_{12,k}^{s}}+\frac{1}{2} m_{12,k}^{s} (\omega_{s,k})^2 \lvert{Q}_{s,k} \rvert^2
\biggr\} + \notag\\
\label{FullHamiltonian}
&+&\frac{2i\alpha}{N} \sum_{k_1,k_2,k_3} \bigl\{W_{1,2,3}^{(1)} {Q}_{+,{k_1}}{Q}_{+,{k_2}}{Q}_{+,{k_3}} +
W_{1,2,3}^{(2)} {Q}_{-,k_1}{Q}_{-,k_2}{Q}_{-,k_3}\\ 
&+&
W_{{1,2,3}}^{(3)} {Q}_{+,k_1}{Q}_{-,k_2}{Q}_{-,k_3} +
W_{1,2,3}^{(4)} {Q}_{-,k_1}{Q}_{+,k_2}{Q}_{+,k_3}
\bigr\}\delta_{k_1+k_2+k_3,0}
\notag.
		\end{eqnarray}
\textcolor{black}{where $P_{\pm,k}=m_{12,k}^\pm \dot{Q}_{\pm,k}$.}

As illustrated in~\eqref{FullHamiltonian}, cubic nonlinearity generates the interaction of three Fourier modes. Possible combinations include acoustical-acoustical-acoustical term, acoustical-acoustical-optical term, acoustical-optical-optical term and optical-optical-optical term. Coefficients of these cubic-interaction terms are given in Appendix. We will show in Section~\ref{s:recurrence due to resonance} that, for the purpose of studying resonances, the acoustical-acoustical-optical term is most relevant.

\section{The classical FPUT recurrence for the diatomic chain}
\label{s:fput recurrence}
We recall that in the original work of Fermi and collaborators, recurrence in the monoatomic chain was observed when the simulation was initialized with the lowest Fourier mode; we will refer to this type of recurrence as the classical FPUT recurrence.

\begin{figure}[htbp]
\centerline{
\includegraphics[width=0.49\textwidth]{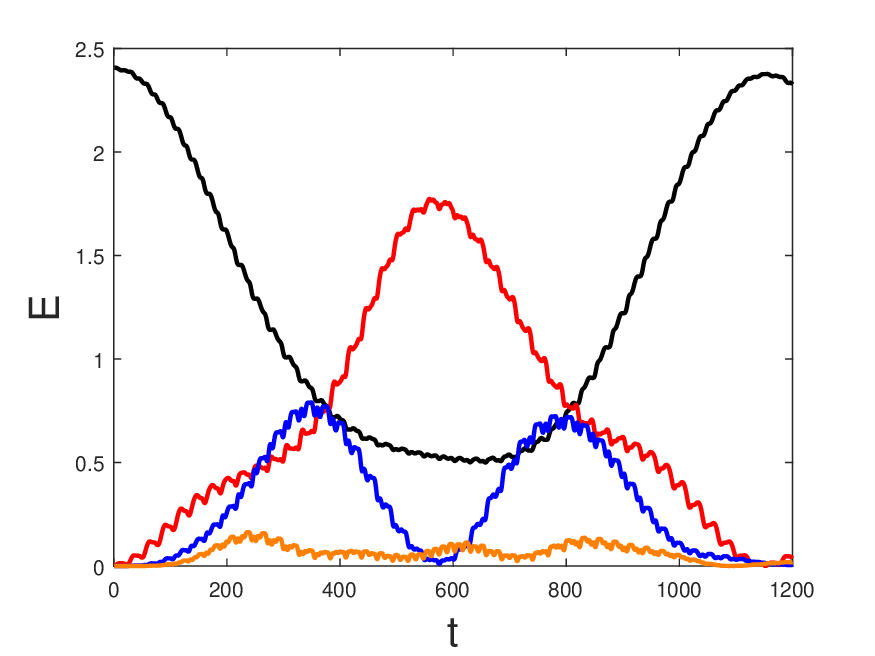}}
\caption{FPUT recurrence in diatomic lattice. The energy densities for four acoustical Fourier modes with lowest frequencies are represented by black, red, blue and orange curves respectively (frequencies from low to high).}
\label{f:fput_recurrence}
\end{figure}

\begin{figure}[b!]
\centerline{
\begin{tikzpicture}
[x=20cm,y=20cm]
  \draw (-1.05,3.16) -- (-0.65,3.16) -- (-0.65,2.97) -- (-1.05,2.97) -- cycle;   
\fill [black]   (-1,3.1489) circle(0.003);
\fill [black]   (-0.9586,3.1287) circle(0.003);
\fill [black]   (-0.9208,3.1099) circle(0.003);
\fill [black]   (-0.8861,3.0941) circle(0.003);
\fill [black]   (-0.8539,3.0781) circle(0.003);
\fill [black]   (-0.8239,3.0626) circle(0.003);
\fill [black]   (-0.7959,3.0492) circle(0.003);
\fill [black]   (-0.7696,3.0362) circle(0.003);
\fill [black]   (-0.7447,3.0237) circle(0.003);
\fill [black]   (-0.7212,3.0107) circle(0.003);
\fill [black]   (-0.6990,2.9974) circle(0.003);

\draw[opacity=1, variable=\x, samples at={-1.009,-1.008,...,-0.69}] plot (\x,-0.49788*\x+2.652);

\draw [gray,dotted] (-1,2.97) -- (-1,3.16);
\node at (-1,2.97) [below] {\scriptsize{$\alpha=0.1$}};

\draw [gray,dotted] (-0.9208,2.97) -- (-0.9208,3.16);

\draw [gray,dotted] (-0.8539,2.97) -- (-0.8539,3.16);
\node at (-0.8539,2.97) [below] {\scriptsize{$\alpha=0.14$}};

\draw [gray,dotted] (-0.7959,2.97) -- (-0.7959,3.16);

\draw [gray,dotted] (-0.7447,2.97) -- (-0.7447,3.16);
\node at (-0.7447,2.97) [below] {\scriptsize{$\alpha=0.18$}};

\draw [gray,dotted] (-0.6990,2.97) -- (-0.6990,3.16);

\draw[gray,dotted] (-1.05,3) -- (-0.65,3);
\node at (-1.05,3) [left] {\scriptsize{$T=1000$}};

\draw[gray,dotted] (-1.05,3.0414) -- (-0.65,3.0414);
\node at (-1.05,3.0414) [left] {\scriptsize{$T=1100$}};

\draw[gray,dotted] (-1.05,3.0792) -- (-0.65,3.0792);
\node at (-1.05,3.0792) [left] {\scriptsize{$T=1200$}};

\draw[gray,dotted] (-1.05,3.1139) -- (-0.65,3.1139);
\node at (-1.05,3.1139) [left] {\scriptsize{$T=1300$}};

\draw[gray,dotted] (-1.05, 3.1461) -- (-0.65, 3.1461);
\node at (-1.05, 3.1461) [left] {\scriptsize{$T=1400$}};

\draw node at (-1.14,3.06) [rotate=90] {\large{$\log_{10}(\mathrm{T})$}};
\draw node at (-0.85,2.95) [below] {\large{$\log_{10}\alpha$}};

\draw node at (-0.805,3.13)  {\large{$\log_{10}\mathrm{T}=-0.5\log_{10}\alpha+2.65$}
}
;

\end{tikzpicture}
}
\caption{Log-log plot of recurrence period vs nonlinear strength $\alpha$. }
\label{f:recurrence_alpha}
\end{figure}
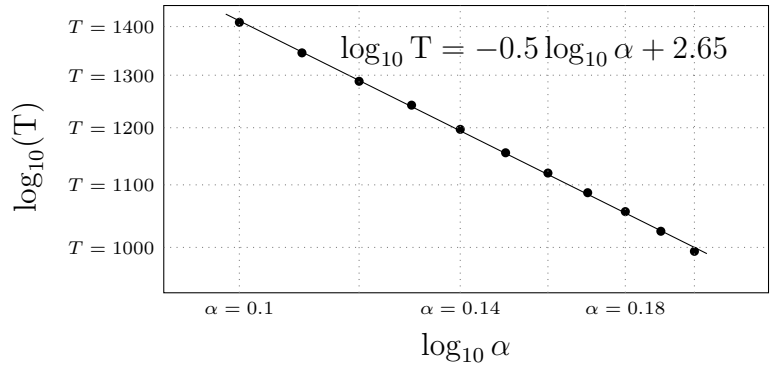

In this Section, we will show numerical simulations of the diatomic lattice in the regime just described. We employ a symplectic integrator in the form of a velocity Verlet algorithm, which is convenient for studying chains of particles characterized by an Hamiltonian formulation. 
We consider a chain with $2N=32$, $\kappa=1$, $\alpha=0.15$, $m_2=5$, and $m_1=1$. As an initial condition, we excite the first acoustical mode; consequently, the initial displacement of each particle is given by the real part of~\eqref{e:transform} with $Q_-(q_1,t)=\mathrm{constant}$, while all the other $Q_\pm$ are zero. More specifically, We consider the following explicit form of the initial data:
\begin{subequations}
\begin{gather}
y_{2n}(0)= \mathrm{C}\cos(2nq_1),\\
y_{2n+1}(0)=\mathrm{C}\beta_-(q_1)\cos((2n+1)q_1),
\end{gather}
\label{e:FPUT_initial}%
\end{subequations}
where $\mathrm{C}$ is  constant, and for our simulations $\mathrm{C}=1$. The initial velocity of all the particles are set to $0$, as in the original FPUT simulation~\cite{fermi1955}.
In Figure~\ref{f:fput_recurrence}, we show the energy density for four acoustical Fourier modes with lowest frequencies: a clear  recurrent behaviour, similar to the one in the monoatomic chain, is observed.
Note that the FPUT recurrence is a result of three-wave non-resonant interactions; such recurrence does not require any specific value of the mass ratio between light and heavy particles. For various values of mass ratio, we observe recursive behaviour of Fourier modes which is similar to that reported in Figure~\ref{f:fput_recurrence}.

The period of the FPUT recurrence is directly related to the strength of nonlinearity $\alpha$ of the system. In Figure~\ref{f:recurrence_alpha}, we show a log-log plot of period vs. the nonlinear strength $\alpha$. For our simulations, $\alpha\in[0.1,0.2]$ and the rest of the system parameters are the same as those in Figure~\ref{f:fput_recurrence}. As illustrated in Figure~\ref{f:recurrence_alpha},
the period of recurrence is proportional to $1/\sqrt{\alpha}$, which is similar to the monatomic case~\cite{TodaBook}.

\begin{figure}[htbp]
\centerline{
\includegraphics[width=0.49\textwidth]{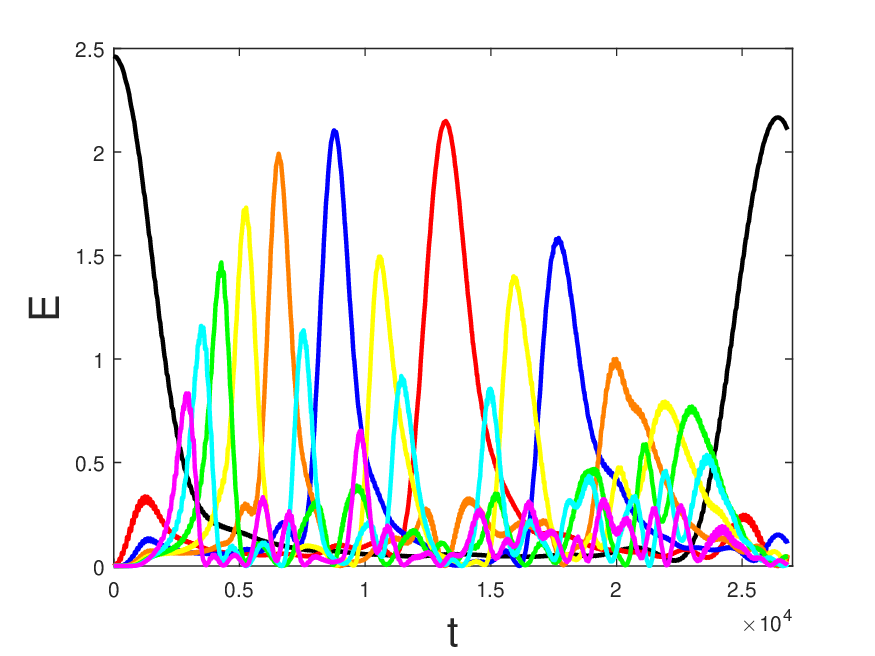}}
\caption{FPUT recurrence in diatomic lattice. The energy densities for eight acoustical Fourier modes with lowest frequencies are represented by black, red, blue, orange, yellow, green, cyan and magenta curves respectively (frequencies from low to high).}
\label{f:fput_recurrence_largeN}
\end{figure}

{\color{black} Figure~\ref{f:fput_recurrence_largeN} shows another example of recurrence for a lattice with 2N=128. As in the classical FPUT problem, the recurrence may involve the exchange of energy among several modes. In contrast, for the second type of recurrence discussed in this work, the dynamics consistently involve only three resonantly interacting modes, as it will be shown in the next Section. } 

\section{Recurrence due to three-wave resonant interaction}
\label{s:recurrence due to resonance}

\begin{figure}[b!]
\centerline{
\includegraphics[width=0.49\textwidth]{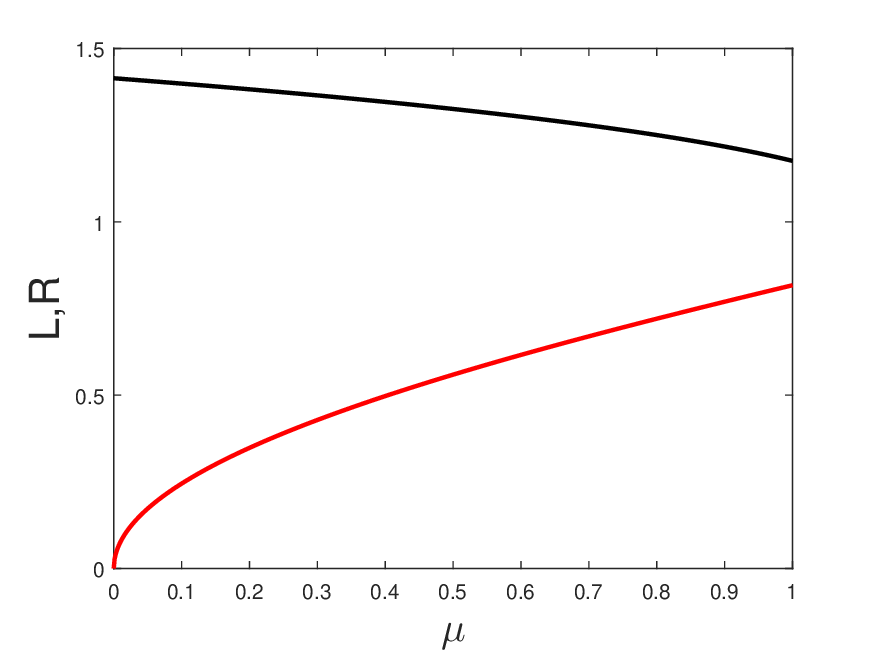}
\includegraphics[width=0.49\textwidth]{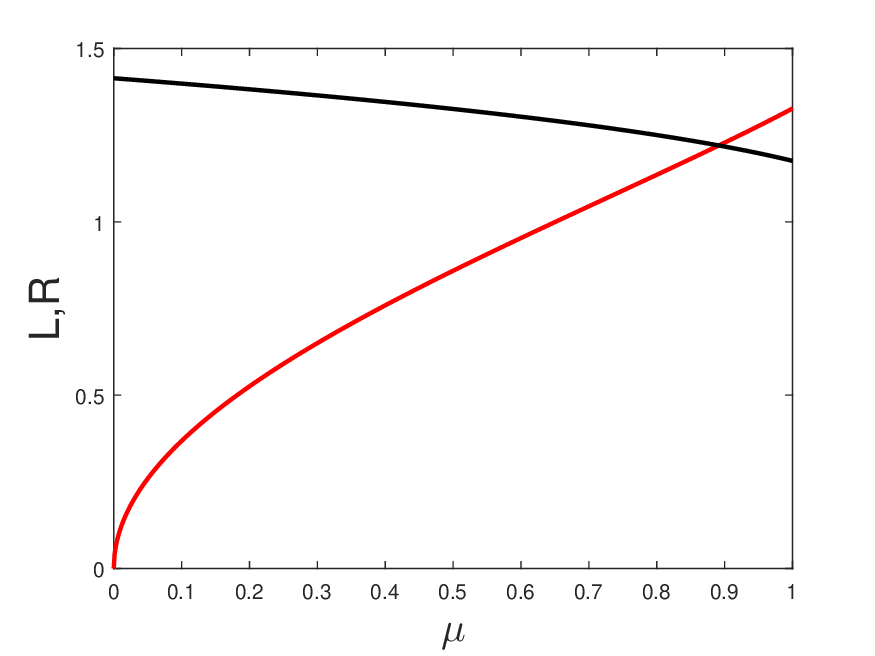}}
\caption{In a diatomic chain with $2N=32$ particles, $L(k_1,k_2,\mu)$ (red curve) and $R(k_3,\mu)$ (black curve) as a function of $\mu$. In the left panel, $k_1=2$, $k_2=4$ and $k_3=6$; in the right panel, $k_1=4$, $k_2=6$ and $k_3=10$.
}
\label{f:resonance}
\end{figure}

Here, we will show that, besides the classical FPUT recurrence, there exists another type of recurrence in diatomic lattice, which is a result of resonant interaction of three Fourier modes. This type of recurrence cannot be found in the monatomic lattice, as three Fourier modes cannot satisfy the resonant condition in monatomic lattice.
\subsection{Existence of three-wave resonance in diatomic FPUT lattice}
\label{ss:resonant}
Here below, we demonstrate theoretically that, for three Fourier modes with acoustical-acoustical-optical combination,
there is a unique mass ratio at which three-wave resonant interaction among these Fourier modes takes place. For acoustical-acoustical-optical combination with arbitrary wave numbers, we also give explicit inequality to predict whether such resonant interaction can be realized or not. 

Fourier modes with acoustical-acoustical-optical combination can have resonant interactions  when they satisfy the following conditions
\begin{gather}
\label{e:wavenumber}
k_1+k_2 = k_3\,,\\
\label{e:frequency}
\omega_{-,k_1}+\omega_{-,k_2}=\omega_{+,k_3}\,.
\end{gather}
The left hand side of~\eqref{e:frequency} can be denoted as
\begin{align}
L(k_1/N,k_2/N,\mu) = \sqrt{1-\sqrt{1-\mu\sin^2\bigg(\frac{\pi k_1}{N}\bigg)}}+\sqrt{1-\sqrt{1-\mu\sin^2\bigg(\frac{\pi k_2}{N}\bigg)}}\,,
\end{align}
and the right hand side of~\eqref{e:frequency} can be denoted as 
\begin{align}
R(k_3/N,\mu) = \sqrt{1+\sqrt{1-\mu\sin^2\bigg(\frac{\pi k_3}{N}\bigg)}}\,,
\end{align}
with 
\begin{align}
    \mu = \frac{4m_1m_2}{(m_1+m_2)^2}.\,
\end{align}
 Noting that $0<4m_1m_2\leq(m_1+m_2)^2$, we have $0<\mu\leq1$. For fixed values of $k_1/N$, $k_2/N$ and $k_3/N$, both $L(k_1/N,k_2/N,\mu)$ and $R(k_3/N,\mu)$ can be seen as a function of $\mu$ defined on domain $(0,1]$.

We have $L(k_1/N,k_2/N,0)=0$ and $R(k_3/N,0)=\sqrt{2}$. Note that $L(k_1/N,k_2/N,\mu)$ is a monotonic increasing function with respect to $\mu$ and that $R(k_3/N,\mu)$ is a monotonic decreasing function with respect to $\mu$. To determine whether an acoustical-acoustical-optical combination with wave numbers satisfying ~\eqref{e:wavenumber} can have resonant interactions or not, we only need to compare $L(k_1/N,k_2/N,1)$ with $R(k_3/N,1)$. If $L(k_1/N,k_2/N,1)>R(k_3/N,1)$, $L(k_1/N,k_2/N,\mu)$ and $R(k_3/N,\mu)$ have one intersection in interval $(0,1]$, at which~\eqref{e:frequency} is satisfied; thus resonant interaction among these modes can be realized. On contrast, if $L(k_1/N,k_2/N,1)< R(k_3/N,1)$, $L(k_1/N,k_2/N,\mu)$ does not intersect with $R(k_3/N,\mu)$ in interval $(0,1]$, which suggests that~\eqref{e:frequency} cannot be satisfied; thus resonant interaction among these modes cannot be realized. 

We consider a diatomic chain with $2N=32$ particles. As shown in the left panel of Figure~\ref{f:resonance}, for $k_1/N=1/8$ and $k_2/N=1/4$, we have $L(1/8,1/4,1)<R(3/8,1)$ and $L(1/8,1/4,\mu)$ does not intersect with $R(3/8,\mu)$, therefore there is no resonant interaction among acoustical-acoustical-optical combination  with wave numbers $2,4,6$. As shown in the right panel of Figure~\ref{f:resonance}, for $k_1/N=1/4$ and $k_2/N=3/8$, we have $L(1/4,3/8,1)>R(5/8,1)$ and $L(1/4,3/8,\mu)$ intersects with $R(5/8,\mu)$ at $\mu=0.8914$, therefore there exists resonant interaction among acoustical-acoustical-optical combination with wave numbers $4,6,10$. 

It is illustrated in~\cite{Pezzi2021} that the acoustical-acoustical-optical resonant interaction can be realized in diatomic lattice if and only if the mass ratio is such that $1<m_2/m_1\leq3$.

\subsection{Recursive behavior due to three-wave resonance in diatomic FPUT lattice: numerical simulations}
\label{ss:recursive}
In this Section, we illustrate that acoustical-acoustical-optical resonant interaction leads to recursive behavior in diatomic FPUT lattice. As shown in Section~\ref{ss:resonant}, there exists a resonant interaction between the acoustical modes $k=4,6$ and the optical mode $k=10$. Solving~\eqref{e:frequency}, we obtain that mass ratio which leads to resonant interaction is $m_2/m_1\approx1.9831$. 
\begin{figure}[htbp]
\centerline{
\includegraphics[width=0.49\textwidth]{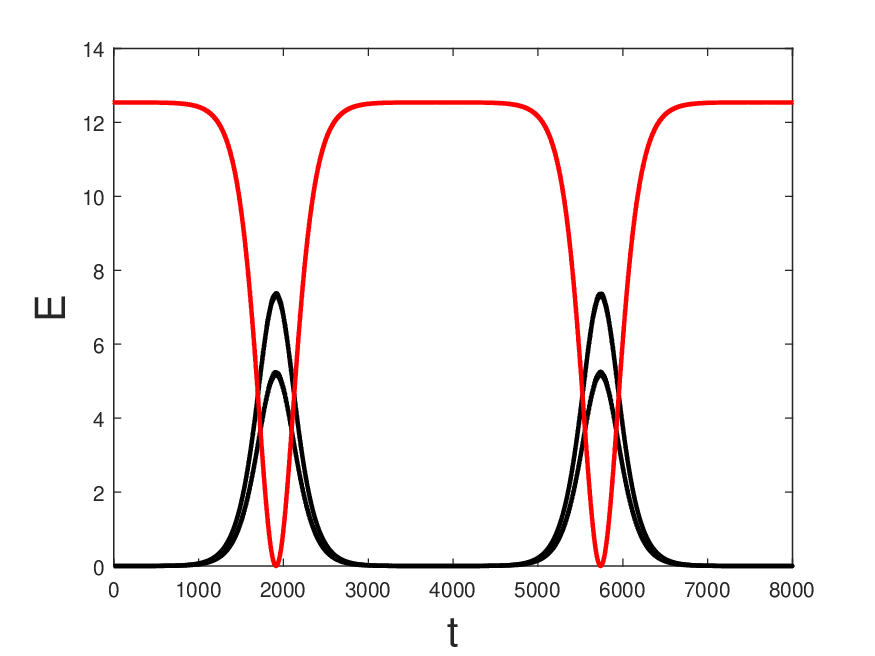}
\includegraphics[width=0.49\textwidth]{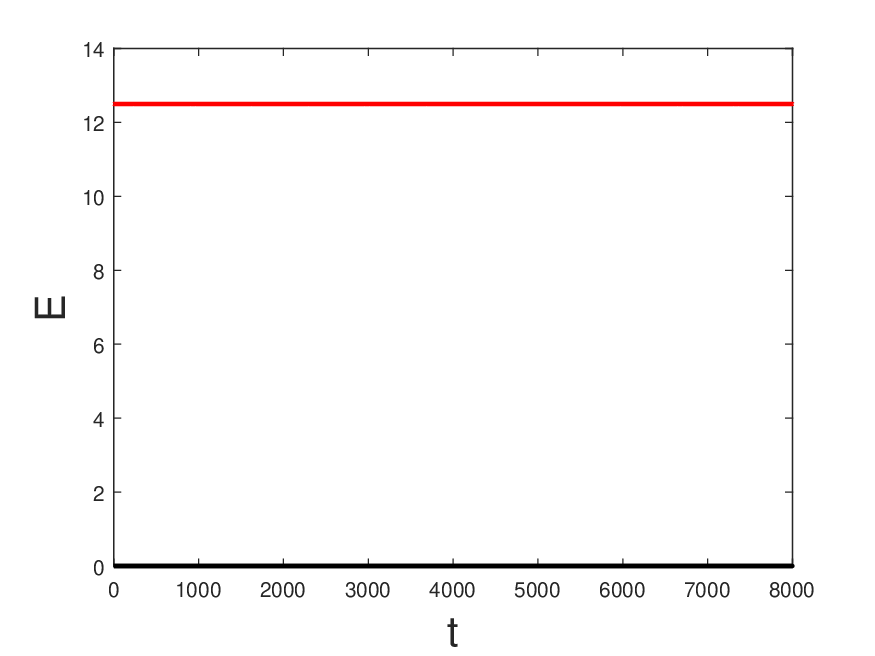}}
\caption{Energy densities of two acoustical modes (black curves) and of optical mode (red curve) as a function of time for a mass ratio of
$m_2/m_1=1.9831$; in the right panel, the same plot in non-resonant conditions for mass ratio $m_2/m_1=5$.}
\label{f:recurrence}
\end{figure}
We simulate the dynamics of a diatomic FPUT lattice with $2N=32$ particles at resonant and non-resonant mass ratio $m_2/m_1=5$, respectively. In both cases, we excite initially the diatomic lattice with optical mode $k=10$. The total time of the simulation is $t=8000$, the time step is $\Delta t=0.01$, the coefficient of linear term is $\kappa=1$, the coefficient of nonlinear term is $\mu=0.015$, and the mass of the light particles is fixed to be $1$.
The initial displacement of each particle is given by the real part of~\eqref{e:transform} with $Q_+(q_{10},t)=\mathrm{constant}$ while all the other $Q_\pm$ are zero, and it can be written in the following explicit form,
\begin{subequations}
\begin{gather}
y_{2n}(0)= \mathrm{C}\cos(2nq_{10}),\\
y_{2n+1}(0)=\mathrm{C}\beta_+(q_{10})\cos((2n+1)q_{10}),
\end{gather}
\label{e:resonant_initial}%
\end{subequations}
where $\mathrm{C}$ is a constant that is set to 1 in the simulations. The initial velocity of all the particles are set to $0$ as the original FPUT simulation~\cite{fermi1955}.
As shown in Figure~\ref{f:recurrence}, at resonant mass ratio we observe recursive behavior of Fourier modes, while at non-resonant mass ratio the energy remains almost constant for all the modes.

\begin{figure}[htbp]
\centerline{
\includegraphics[width=0.95\textwidth]{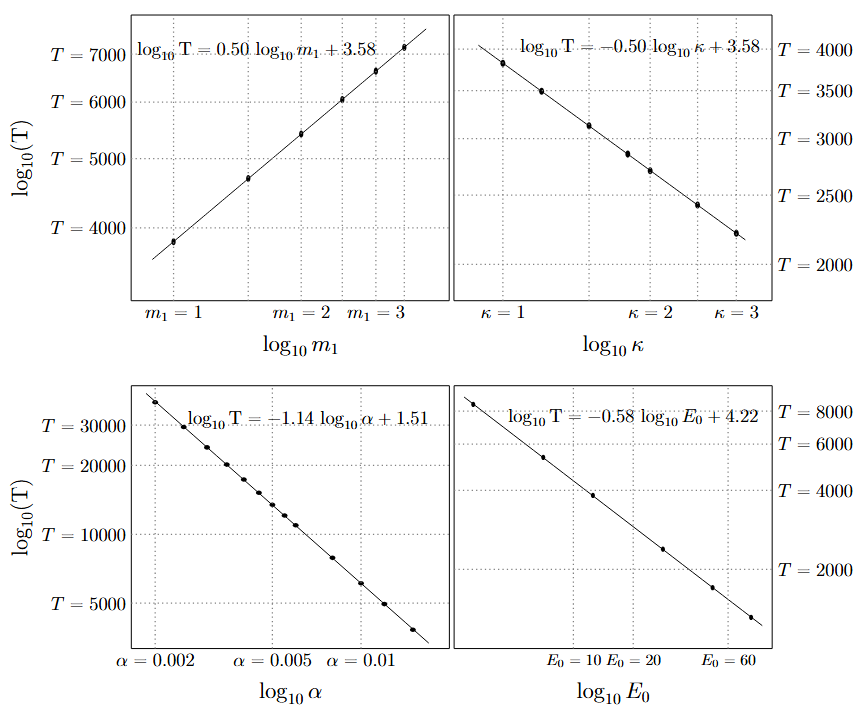}
}
\caption{Log-log plot of period for recurrence with respect to the mass of light particle $m_1$, 
the coefficient of linear term $\kappa$, the coefficient of nonlinear term $\alpha$ and the energy $E_0$ initially given to the optical Fourier mode. 
}
\label{f:fitting}
\end{figure}

Next we determine the dependence of the recurrence period on system parameters, including the light mass $m_1$\footnote{To ensure that the recurrence occurs, we need to fix the mass ratio. Therefore once $m_1$ is given, $m_2$ is determined.}, the coefficient of linear term $\kappa$, the coefficient of nonlinear term $\alpha$ and the energy initially given to the optical Fourier mode $E_0$. For each of the parameters, by changing only the relevant one and fixing the other ones, we measure the period at various values of system parameters. 
We show log-log plot of period with respect to $m_1$, $\kappa$, $\alpha$ and $E_0$ in Figure~\ref{f:fitting}. 

\begin{figure}[htbp]
\centerline{
\includegraphics[width=0.49\textwidth]{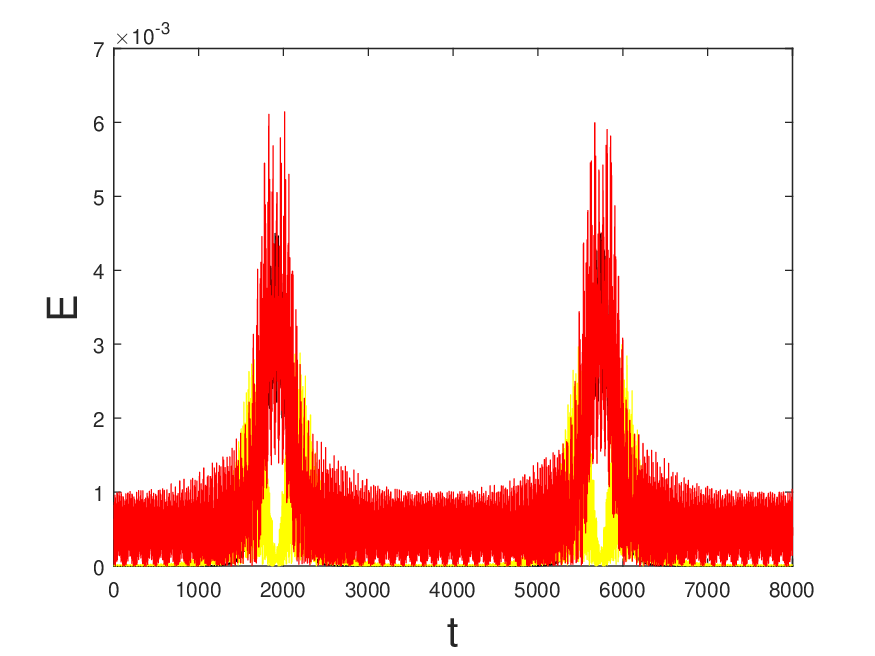}
}
\caption{Energy as a function of time for Fourier modes not involved in the resonant interaction. The system parameters are the same as those used in the left panel of Figure~\ref{f:recurrence}.}
\label{f:energy}
\end{figure}

Note that the recursive behavior studied in this section differs from the FPUT recurrence discussed in Section~\ref{s:fput recurrence} in several important aspects. The FPUT recurrence arises purely from nonlinearity and can therefore be observed for arbitrary mass ratios, provided that the coefficient $\alpha$ of the nonlinear term is sufficiently large.
As illustrated in Figure~\ref{f:fput_recurrence}, FPUT recurrence is observed in a diatomic lattice with mass ratio $m_2/m_1=5$, for which resonance-induced recurrence is not possible. Recall that three-wave resonance can occur only when $1 < m_2/m_1 \leq 3$.
In contrast, the recursive behavior studied in this section results from the combined effects of nonlinearity and resonant interactions among three Fourier modes. For a fixed wavenumber of the initially excited optical mode, this type of recursive behavior can occur only for specific mass ratios satisfying the resonance conditions.
Another important distinction is that, whereas classical FPUT recurrences may involve the participation of many modes, the recurrence discussed here is essentially confined to three Fourier modes, with the vast majority of the system's energy remaining concentrated within this resonant triad. For the recursive behavior shown in the left panel of Figure~\ref{f:recurrence}, the maximum energy contained in non-recursive Fourier modes is less than $7\times10^{-3}$, as illustrated in Figure~\ref{f:energy}, while the total energy of the system is equal to $12.5379$. 

\begin{figure}[b!]
\centerline{
\includegraphics[width=0.49\textwidth]{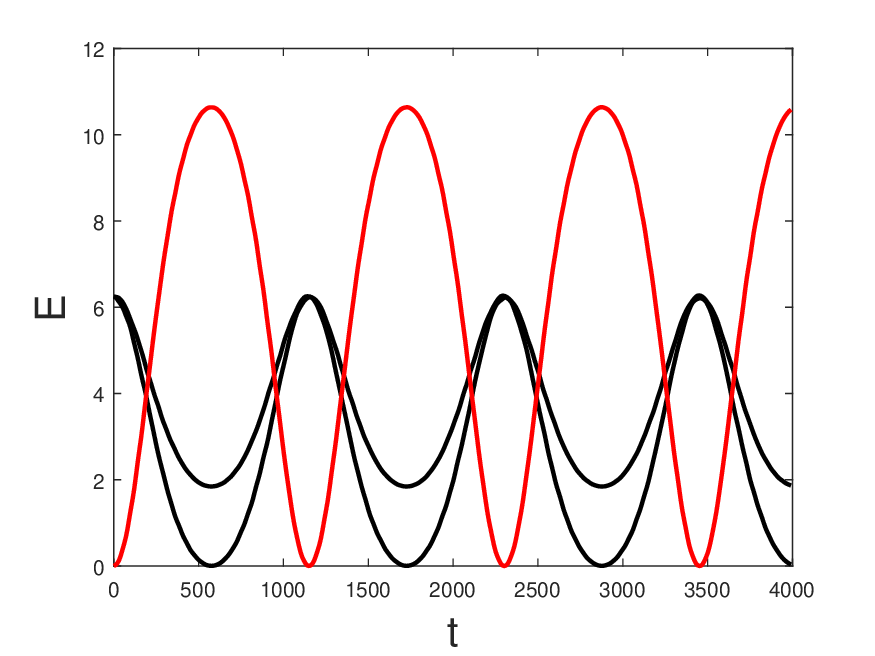}
\includegraphics[width=0.49\textwidth]{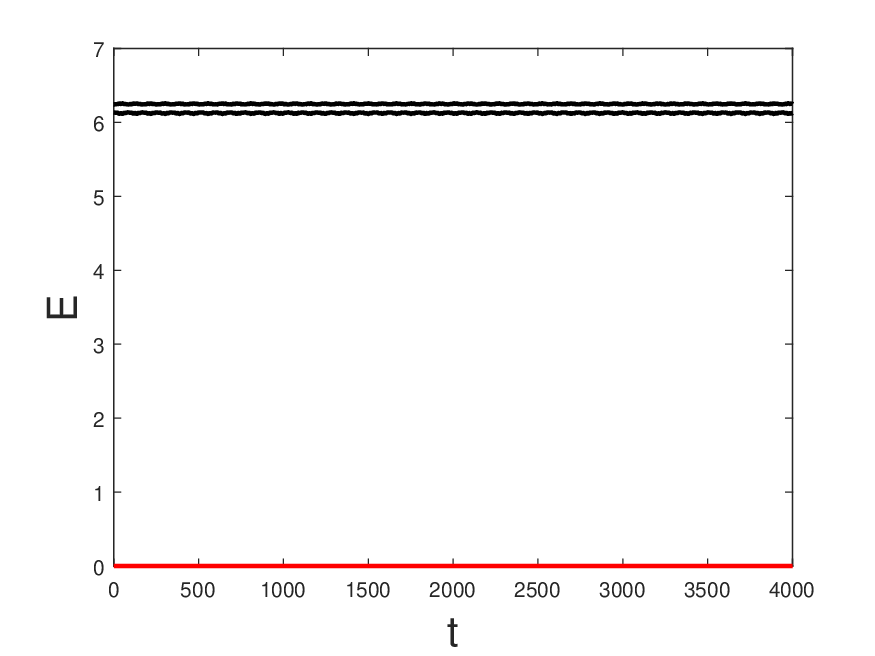}}
\caption{Energy of the two acoustical modes (black curves) and the optical mode (red curve) as functions of time. In our simulations, the two acoustical modes are initially excited. In the left panel, the mass ratio is fixed at the resonant value $m_2/m_1=1.9831$, while in the right panel the mass ratio is fixed at the non-resonant value $m_2/m_1=5$.}
\label{f:recurrence_acoustical}
\end{figure}

It is also worth noting that recursive behavior can be observed at the mass ratio $m_2/m_1=1.9831$ by initially exciting the acoustical modes $k=4$ and $k=6$. With the same system parameters as before, the explicit form of the initial condition is given by 
\begin{subequations}
\begin{gather}
y_{2n}(0)= \mathrm{C}_1\cos(2nq_{4})+\mathrm{C}_2\cos(2nq_{6}),\\
y_{2n+1}(0)=\mathrm{C}_1\beta_-(q_{4})\cos((2n+1)q_{4})+\mathrm{C}_2\beta_-(q_{6})\cos((2n+1)q_{6}),
\end{gather}
\label{e:resonant_initial_1}%
\end{subequations}
with $\dot{y}_n(0)=0$ for all $n$, where $\mathrm{C}_1$ and $\mathrm{C}_2$ are arbitrary constants. 
We show the corresponding recursive behavior of the system for the mass ratio $m_2/m_1=1.9831$ in the left panel of Figure~\ref{f:recurrence_acoustical}. For comparison, we also excite the system using the initial condition~\eqref{e:resonant_initial_1} with the non-resonant mass ratio $m_2/m_1=5$, and present the resulting dynamics in the right panel of Figure~\ref{f:recurrence_acoustical}. 

In our simulations, we set $C_1=C_2=0.5319$ so that the total energy of the system matches that of the case in which a single optical mode is initially excited. Note that, for the same total system energy, the recursive behavior shown in Figure~\ref{f:recurrence_acoustical} occurs on a much faster time scale than the behavior shown in Figure~\ref{f:recurrence}. We explain the underlying mechanism by analyzing the structure of a reduced model in Fourier space in Section~\ref{sss:Fourier}. 

\section{The three-mode system and its recursive solution }
\label{sss:Fourier}

As previously shown, for recurrence induced by resonant interactions, the majority of the system energy remains confined to the Fourier modes participating in the resonance, while the energy carried by the remaining Fourier modes is negligibly small. In this section, we consider a reduced model in Fourier space that includes only the two acoustical modes and the single optical mode involved in the resonant interaction. We will show that this reduced model is able to reproduce the main dynamics of the full system.

The Hamiltonian equations for canonical Fourier modes are given by
\begin{equation}
\label{e:Hamiltoneq}
\textcolor{black}{
\dot{Q}_{\pm,k}=N\frac{\partial H}{\partial P_{\pm,k}^{*}},\qquad
\dot{P}_{\pm,k}=-N\frac{\partial H}{\partial Q_{\pm,k}^{*}}
}
\end{equation}

Recalling~\eqref{FullHamiltonian}, we have 
\begin{subequations}
\label{nlsystemdiagonal}
\begin{eqnarray}
\ddot{Q}_{+,1}+(\omega^+_{1})^2{Q}_{+,1}&=&
\frac{2i\alpha}{N m^+_{12,1}}\sum_{2,3}\bigl\{V_{1,2,3}^{(1)}Q_{+,2}{Q}_{+,3}+
V_{1,2,3}^{(2)}{Q}_{-,2}{Q}_{-,3}+
V_{1,2,3}^{(3)}{Q}_{+,2}{Q}_{-,3}
\bigr\} \delta_{1,2+3}, \label{nlsystemdiagonal1}\\ 
\ddot{Q}_{-,1}+(\omega^-_{1})^2{Q}_{-,1}&=&
\frac{2i \alpha}{N m^-_{12,1}}\sum_{2,3}
\bigl\{T_{1,2,3}^{(1)}{Q}_{+,2}{Q}_{+,3}+
T_{1,2,3}^{(2)}{Q}_{-,2}{Q}_{-,3}+
T_{1,2,-3}^{(3)}{Q}_{+,2}{Q}_{-,3}\bigr\} \delta_{1,2+3}.
\label{nlsystemdiagonal2}
\end{eqnarray}
\end{subequations}
For purposes of convenience, in~\eqref{nlsystemdiagonal} and hereafter we set $Q_{\pm,k_i}=Q_{\pm,i}$, $\omega_{\pm,k_i}=\omega^{\pm}_{i}$, $m^{\pm}_{12,k_i}=m^{\pm}_{12,i}$ and $\delta_{k_1,k_2+k_3}=\delta_{1,2+3}$. The coefficients in~\eqref{nlsystemdiagonal} are reported in the Appendix.
Introducing the normal variables
\begin{equation}
a_k^{\pm}=\frac{i}{\sqrt{2m^{\pm}_{12,k}\omega_{\pm,k}}}
(P_{\pm,k}-im^{\pm}_{12,k}\omega_{\pm,k}Q_{\pm,k}),
\label{e:expression_a}
\end{equation}
we obtain the following equations of motion for $a^{\pm}_k$
\begin{subequations}
\label{eq:3wo}
		\begin{eqnarray}
        \notag
			&i\frac{da_1^{+}}{dt}=\omega_1^{+}a_1^{+} +\sum_{2,3}\big\{ [\bar{V}_{1,2,3}^{(1)} a_2^{+} a_3^{+}+\bar{V}_{1,2,3}^{(2)} a_2^{-} a_3^{-}
			+ \bar{V}_{1,2,3}^{(3)} a_2^{+} a_3^{-}] \delta_{1,2+3}+
			[\bar{V}_{1,-2,-3}^{(1)} a_2^{+*} a_3^{+*}+ \bar{V}_{1,-2,-3}^{(2)} a_2^{-*} a_3^{-*} +
			\\&
		 \bar{V}_{1,-2,-3}^{(3)} a_2^{+*} a_3^{-*}]\delta_{1+2+3,0}+
			[2\bar{V}_{1,2,-3}^{(1)} a_2^{+} a_3^{+*}+2\bar{V}_{1,2,-3}^{(2)} a_2^{-} a_3^{-*}
			+ \bar{V}_{1,2,-3}^{(3)} a_2^{+} a_3^{-*}+\bar{V}_{1,-3,2}^{(3)} a_2^{-} a_3^{+*}] \delta_{1,2-3}\big\},
		\end{eqnarray}	
		\begin{eqnarray}
        \notag
			&i\frac{da_1^{-}}{dt}=\omega_1^{-}a_1^{-} +\sum_{2,3}\big\{ [\bar{T}_{1,2,3}^{(1)} a_2^{+} a_3^{+}+\bar{T}_{1,2,3}^{(2)} a_2^{-} a_3^{-}
			+ \bar{T}_{1,2,3}^{(3)} a_2^{+} a_3^{-}] \delta_{1,2+3}+
			[\bar{T}_{1,-2,-3}^{(1)} a_2^{+*} a_3^{+*}+ \bar{T}_{1,-2,-3}^{(2)} a_2^{-*} a_3^{-*} +
			\\& 
			\bar{T}_{1,-2,-3}^{(3)} a_2^{+*} a_3^{-*}]\delta_{1+2+3,0}+
			[2\bar{T}_{1,2,-3}^{(1)} a_2^{+} a_3^{+*}+2\bar{T}_{1,2,-3}^{(2)} a_2^{-} a_3^{-*}
			+ \bar{T}_{1,2,-3}^{(3)} a_2^{+} a_3^{-*}+\bar{T}_{1,-3,2}^{(3)} a_2^{-} a_3^{+*}] \delta_{1,2-3}\big\}.
		\end{eqnarray}	
	\end{subequations}
The coefficients in~\eqref{eq:3wo} are reported in the Appendix.
To be consistent with Section~\ref{s:recurrence due to resonance}, in this section we maintain acoustical modes with wavenumbers $k=4,6$, and optical mode with wavenumber $k=10$, and set all the other modes to be $0$. The time evolution equations of these three Fourier modes in the reduced model are given by 
\begin{equation}
\begin{split}
&i\frac{d a_{10}^+}{d t}=\omega_{10}^{+}a_{10}^++2\widetilde{V}^{+--}_{10,4,6}a_4^-a_6^-,\\
&i\frac{d a_{6}^-}{d t}=\omega_{6}^{-} a_6^-+2\widetilde{V}^{+--*}_{10,4,6}a_{10}^+a_4^{-*},\\
&i\frac{d a_{4}^-}{d t}=\omega_{4}^{-} a_4^-+2\widetilde{V}^{+--*}_{10,4,6}a_{10}^+a_6^{-*},
\end{split}
\label{e:motion in Fourier}%
\end{equation}
where
\begin{equation}
\widetilde{V}^{+--}_{10,4,6}=-\frac{i\alpha}{N\sqrt{2m_{12,10}^+m_{12,6}^-m_{12,4}^-\omega_{10}^+\omega_6^-\omega_4^-}}V_{10,4,6}^{+--},
\end{equation}
\begin{equation}
V_{10,4,6}^{+--}=-A_{-10,4,6}^{+--}-A_{4,-10,6}^{-+-}-A_{6,-10,4}^{-+-},
\end{equation}
\begin{equation}
A_{k_1,k_2,k_3}^{s_1,s_2,s_3}=(\beta_{s_1,k_1}+\beta_{s_2,k_2}\beta_{s_3,k_3})\sin(a q_{k_1}),
\end{equation}
and $m_{12}^{\pm}$ is given in~\eqref{e:m12}.
It is straightforward to see that  $\widetilde{V}_{10,4,6}^{+--}$ is a purely imaginary number.  

Introducing $\bar{a}_k^s=a_k^s\mathrm{e}^{i\omega_k^st}$ with $s=\pm$, we write~\eqref{e:motion in Fourier} as 
\begin{equation}
\begin{split}
&i\frac{d \bar{a}_{10}^+}{d t}=2\widetilde{V}^{+--}_{10,4,6}\bar{a}_4^-\bar{a}_6^-\mathrm{e}^{i(\omega_{10}^+-\omega_4^--\omega_6^-)t},\\
&i\frac{d \bar{a}_{6}^-}{d t}=2\widetilde{V}^{+--*}_{10,4,6}\bar{a}_{10}^+\bar{a}_4^{-*}\mathrm{e}^{i(-\omega_{10}^++\omega_4^-+\omega_6^-)t},\\
&i\frac{d \bar{a}_{4}^-}{d t}=2\widetilde{V}^{+--*}_{10,4,6}\bar{a}_{10}^+\bar{a}_6^{-*}\mathrm{e}^{i(-\omega_{10}^++\omega_4^-+\omega_6^-)t}.
\end{split}
\label{e:motion in Fourier_1}%
\end{equation}
Our reduction leading to the three–wave system \eqref{eq:ABC1} implicitly relies on the assumption that the resonant triad $(k_A,k_B,k_C)$ is \emph{dynamically isolated}. Such assumption is true for our model where the number of total particles is relatively small.
In the large–box limit, three–wave resonances typically organize into clusters of interconnected triads, and such assumption in general does not hold anymore. 

\begin{figure}[b!]
\centerline{
\includegraphics[width=0.49\textwidth]{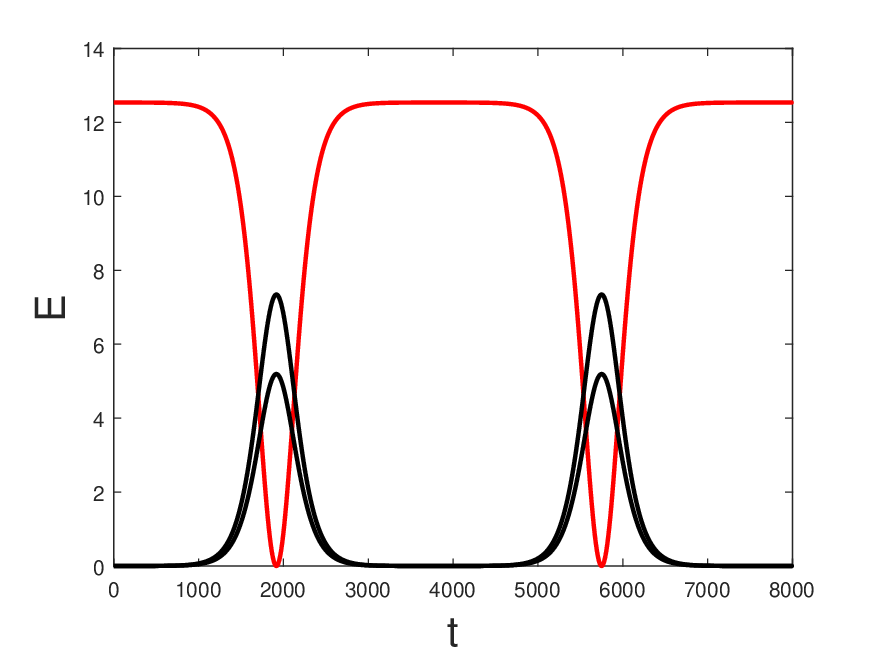}
\includegraphics[width=0.49\textwidth]{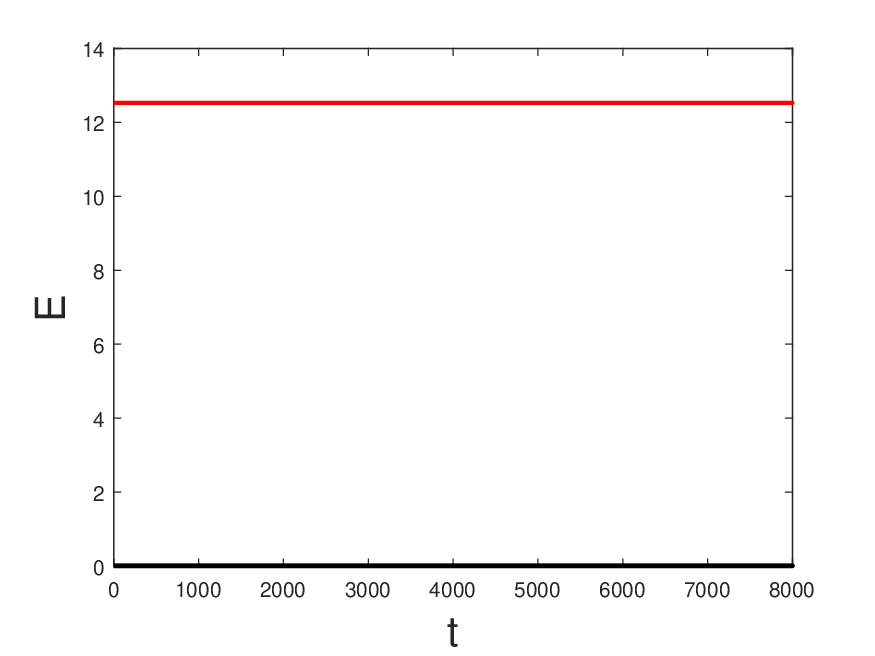}}
\caption{Energy of two acoustical modes (black curves) and of optical mode (red curve) as a function of time in a reduced model characterized by~\eqref{e:motion in Fourier_1} in Fourier space. In the left panel, the lattice is at resonant mass ratio
$m_2/m_1=1.9831$; in the right panel, the lattice is at non-resonant mass ratio $m_2/m_1=5$.}
\label{f:recurrence Fourier}
\end{figure}

We can show that if $(\bar{a}_{10}^+,\bar{a}_{6}^-,\bar{a}_{4}^-)$ is a solution of~\eqref{e:motion in Fourier_1},  $(\bar{a}_{10}^+\mathrm{e}^{\mathrm{i}\theta_1},\bar{a}_{6}^-\mathrm{e}^{\mathrm{i}\theta_2},\bar{a}_{4}^-\mathrm{e}^{\mathrm{i}(\theta_1-\theta_2)})$, $(\bar{a}_{10}^+\mathrm{e}^{\mathrm{i}\theta_1},\bar{a}_{6}^-\mathrm{e}^{\mathrm{i}(\theta_1-\theta_2)},\bar{a}_{4}^-\mathrm{e}^{\mathrm{i}\theta_2)}$ and $(\bar{a}_{10}^+\mathrm{e}^{\mathrm{i}(\theta_1+\theta_2)},\bar{a}_{6}^-\mathrm{e}^{\mathrm{i}\theta_1},\bar{a}_{4}^-\mathrm{e}^{\mathrm{i}\theta_2})$, where $\theta_1,\theta_2$ are arbitrary real numbers, are also solutions of~\eqref{e:motion in Fourier_1}. Therefore, for the purpose of studying recurrence, where initially one of the three Fourier modes is $0$, we can always rescale the initial value of the other two Fourier modes to real numbers without changing the value of $|a_k^{\pm}|^2$. Recall that at resonant mass ratio we have $\omega_{10}^+-\omega_6^--\omega_4^-=0$ and $\widetilde{V}^{+--}_{10,4,6}$ is purely imaginary, thus for real initial values the solutions of~\eqref{e:motion in Fourier_1} will remain real. 
Therefore, at resonant mass ratio for real initial values we can further simplify~\eqref{e:motion in Fourier_1} to  
\begin{equation}
\begin{split}
\frac{d \bar{a}_{10}^+}{d t}=-2\,i\,\widetilde{V}^{+--}_{10,4,6}\bar{a}_4^-\bar{a}_6^-,\\
\frac{d \bar{a}_{6}^-}{d t}=2\,i\,\widetilde{V}^{+--}_{10,4,6}\bar{a}_{10}^+\bar{a}_4^{-},\\
\frac{d \bar{a}_{4}^-}{d t}=2\,i\,\widetilde{V}^{+--}_{10,4,6}\bar{a}_{10}^+  \bar{a}_6^{-}.
\end{split}
\label{e:motion in Fourier_2}
\end{equation}
From~\eqref{e:motion in Fourier_2}, it is straightforward to show that $J=\omega_{10}^+(\bar a_{10}^+)^2+\omega_{6}^-(\bar a_6^-)^2+\omega_{4}^-(\bar a_4^-)^2$ is a constant of motion, together with the two Manley–Rowe invariants $I_{10,6}=(\bar a_{10}^+)^2+(\bar a_6^-)^2$ and $I_{10,4}=(\bar a_{10}^+)^2+(\bar a_4^-)^2$.
Thus~\eqref{e:motion in Fourier_2} is the standard three–wave interaction model for an isolated resonant triad and is known to be a classical integrable
system \cite{bustamante2011resonance}.

\paragraph{Numerical evidence for recurrence in a reduced model} In this Section, we numerically integrate~\eqref{e:motion in Fourier_1} with initial conditions corresponding to those in Section~\ref{ss:recursive}.

To reproduce recurrence reported in Figure~\ref{f:recurrence} in the reduced model, for our simulations at $t=0$ we set $a_{10}^+=2.8944$ so that the optical Fourier mode has the same energy as Figure~\ref{f:recurrence}. It is worth noting that exciting only the optical mode cannot lead to recurrence in the reduced model.  To induce recurrence, we need to give one of the acoustical modes a small perturbation. Hence, for our simulations we set $a_4^-=5.2\times10^{-3}$, which we adjust so that the recurrence in Figure~\ref{f:recurrence Fourier} has the same time scale as that in Figure~\ref{f:recurrence}. For mass ratio $m_2/m_1=1.9831$ we observe recursive behavior as illustrated in the left panel of Figure~\ref{f:recurrence Fourier}. Such recurrence disappears for the same initial condition at mass ratio $m_2/m_1=5$ as illustrated in the right panel of Figure~\ref{f:recurrence Fourier}.

\begin{figure}[htbp]
\centerline{
\includegraphics[width=0.49\textwidth]{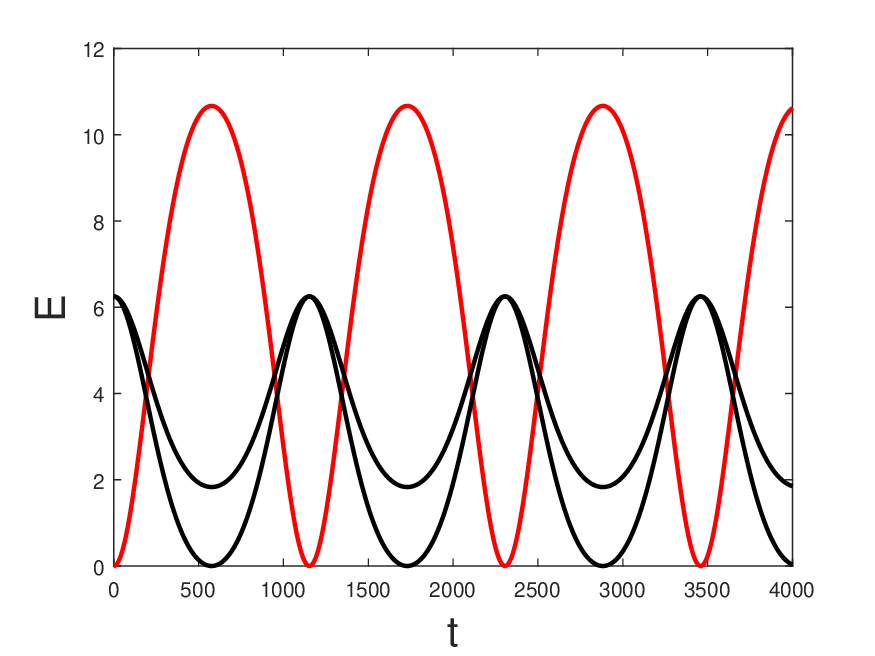}
\includegraphics[width=0.49\textwidth]{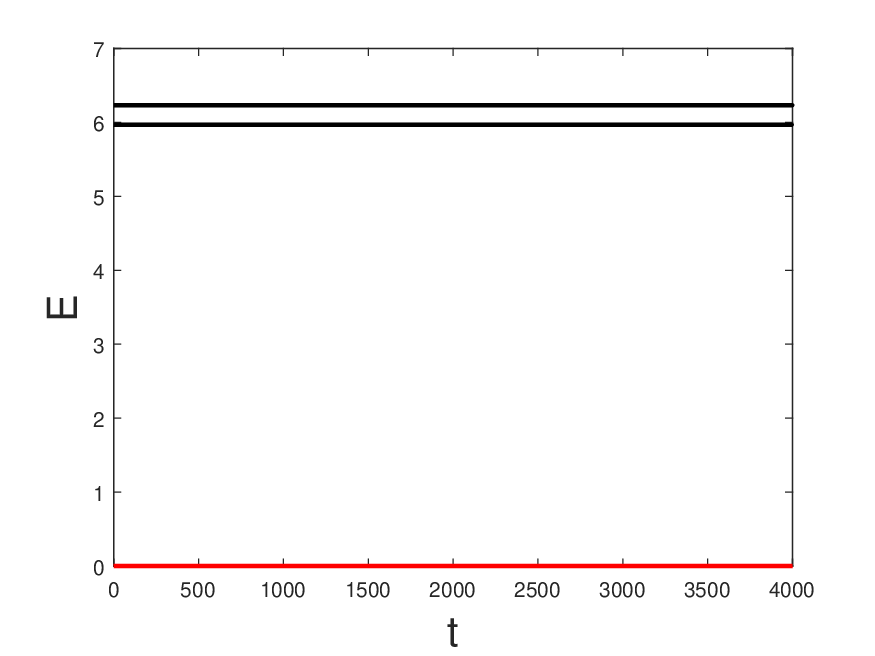}}
\caption{Energy of two acoustical modes (black curves) and of optical mode (red curve) as a function of time in a reduced model characterized by~\eqref{e:motion in Fourier_1} in Fourier space. Initially we excite two acoustical modes. In the left panel, the lattice is at resonant mass ratio
$m_2/m_1=1.9831$; in the right panel, the lattice is at non-resonant mass ratio $m_2/m_1=5$.}
\label{f:recurrence Fourier 1}
\end{figure}

The recurrence reported in Figure~\ref{f:recurrence_acoustical} can be obtained in our reduced model by applying the same initial condition as Section~\ref{ss:recursive}. By initially exciting two acoustical modes at resonant mass ratio $m_2/m_1=1.9831$ in our reduced model, we observe recursive behavior as illustrated in the left panel of Figure~\ref{f:recurrence Fourier 1}. Such recurrence disappears at non-resonant mass ratio $m_2/m_1=5$ as illustrated in the right panel of Figure~\ref{f:recurrence Fourier 1}. 

By analyzing our reduced model, we can explain why the time scale of the recursive behavior reported in Figure~\ref{f:recurrence} is much slower than that of the recurrence reported in Figure~\ref{f:recurrence_acoustical}. The recursive behavior reported in Figure~\ref{f:recurrence} is in fact composed of two stages. In stage one, the involving acoustical modes obtain some small but nonzero energy through non-resonant interactions. And in stage two the acoustical modes interchange energy with the initially excited optical mode through resonant interaction, which leads to recursive behavior. For recurrence in Figure~\ref{f:recurrence} initially only one mode is excited, to induce resonant interaction the other modes need to obtain energy through non-resonant interaction first; while for recurrence in Figure~\ref{f:recurrence_acoustical} initially two modes are excited, hence the resonant interaction can take place immediately.

\section{Linear limit of recursive behavior}
\label{s:linear limit}

With the decrease of the coefficient $\alpha$ of the nonlinear term, the system approaches its linear regime, where recursive behavior does not exist. In this Section, we study the linear limit of both FPUT recurrence and recurrence due to resonance.

\begin{figure}[htbp]
\centerline{
\includegraphics[width=0.49\textwidth]{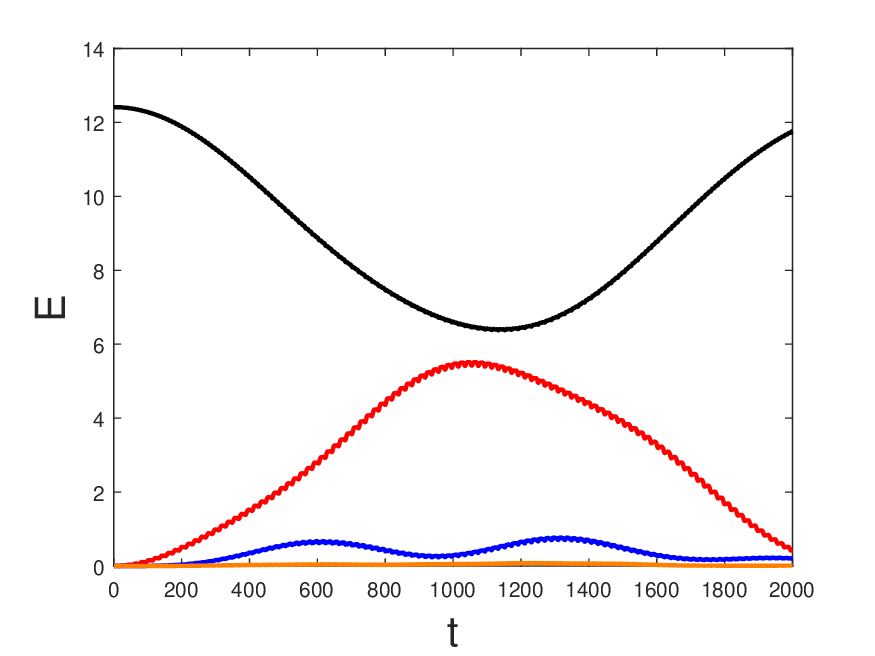}
\includegraphics[width=0.49\textwidth]{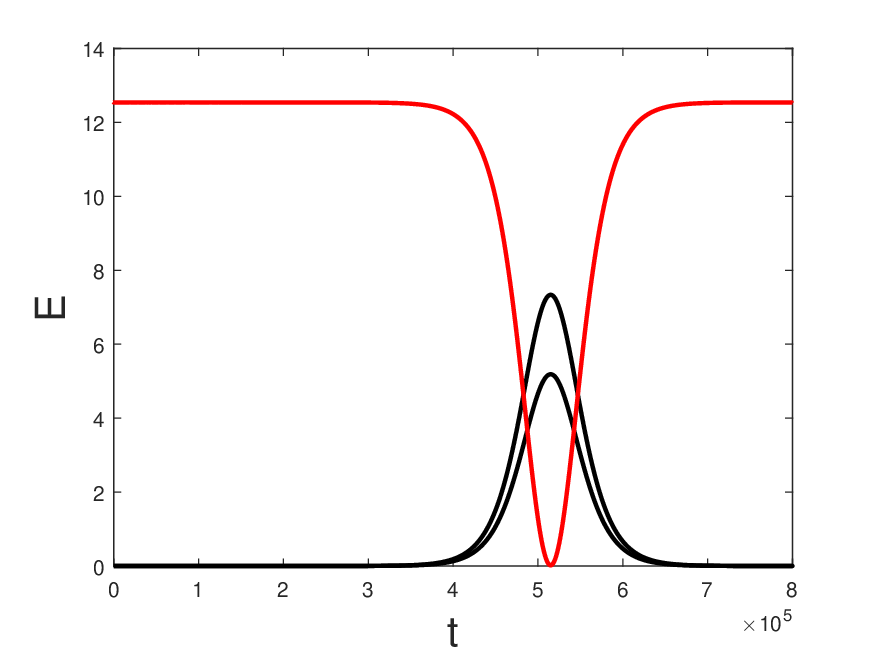}
}
\caption{For $m_1=1$, $m_2=1.9831$ and $\kappa=1$, the FPUT recurrence for $\alpha=0.015$ (left panel); the recurrence due to resonant interactions for $\alpha=0.0001$ (right panel). As shown in the left panel, for FPUT recurrence at relatively small value of $\alpha$, the value of $|E_0-E_{\mathrm{min}}|/E_0$, which quantifies the strength of a recurrence, reduces significantly compared to that in Figure~\ref{f:fput_recurrence}, for which $\alpha$ is relatively large. As shown in the right panel, for recurrence due to resonance, the value of $|E_0-E_{\mathrm{min}}|/E_0$ stays near $1$ even at relatively small value of $\alpha$. The period of recurrence becomes longer at relatively small value of $\alpha$ for recurrence of both types. 
}
\label{f:linear}
\end{figure}

The linear limits of these two types of recurrence are very distinct.
In Figure~\ref{f:linear}, we illustrate recursive behavior of both types for relatively small values of $\alpha$.
The left panel is the FPUT recurrence for $\alpha=0.015$ and the right panel is the recurrence due to resonant interaction between Fourier modes for $\alpha=0.0001$. As shown in the figure, for both types of recurrence, with the decrease of $\alpha$ the period of recursive behavior becomes longer. 
However, the amount of energy exchange among Fourier modes are different in these two types of recurrence. With the decrease of $\alpha$, the exchange of energy decreases for the classical FPUT recurrence, while it stays unchanged for recurrence due to resonance.

To quantify the strength of a recurrence, we introduce the quantity $|E_0-E_{\mathrm{min}}|/E_0$, where $E_0$ represents the energy initially given to the excited Fourier mode and $E_{\mathrm{min}}$ represents the minimum energy in the initially excited Fourier mode. When the majority of $E_0$ flows to other Fourier modes,  $|E_0-E_{\mathrm{min}}|/E_0$ approaches $1$. By contrast, when only a small amount of $E_0$ flows to other Fourier modes, $|E_0-E_{\mathrm{min}}|/E_0$ approaches $0$.  

We simulate a diatomic chain with parameters $N$, $m_2$, $m_1$ and $\kappa$ the same as those in Section~\ref{ss:recursive}. For comparison purposes, the energy initially given to the system for all the trials in this section is the same as that in Section~\ref{s:recurrence due to resonance}, i.e., $E_0=12.5379$. 
To investigate linear limit of recurrence of both types, we initially excite the acoustical mode with $k=1$ and the optical mode with $k=10$ respectively at different values of $\alpha$. For the classical FPUT recurrence, there is a transition region, in which the value of $|E_0-E_{\mathrm{min}}|/E_0$ changes from $0$ to $0.9$, while for recurrence due to resonance the value of $|E_0-E_{\mathrm{min}}|/E_0$ stays close to $1$ at different values of $\alpha$. 
This suggests that, for the classical FPUT recurrence, the exchange of energy between modes decreases as $\alpha$ is reduced; by contrast, for recurrence due to resonance the exchange of energy is independent from $\alpha$ (provided $\alpha>0$). This can be explained by analyzing the structure of~\eqref{e:motion in Fourier_1}: the change of the value of $\alpha$ amounts to rescaling the time, which suggests that we can observe recurrence due to resonance at arbitrary positive value of $\alpha$ by rescaling the time accordingly.

\begin{figure}[htbp]
\centerline{
\includegraphics[width=0.49\textwidth]{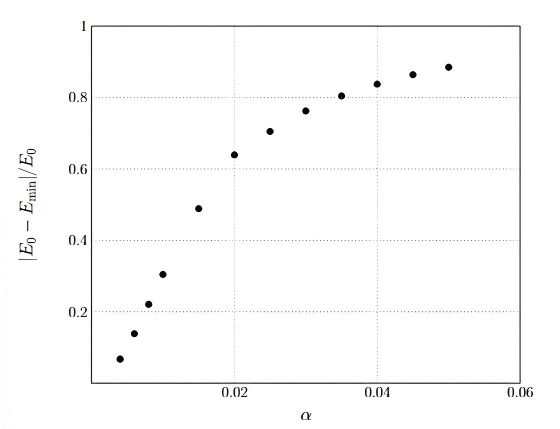}}
\caption{$|E_0-E_{\mathrm{min}}|/E_0$ vs $\alpha$ for the classical FPUT recurrence.
}
\label{f:transition.jpg}
\end{figure}

\section{Robustness of recurrence}
\label{s:variance} 
Here, we test the robustness of the recurrence by incorporating higher-order nonlinear terms into the FPUT lattice and introducing random fluctuations in the particle masses.

\subsection{Effect of higher-order terms on recurrence}
FPUT lattice is a universal model to characterize dynamics of various nonlinear chains in the weakly nonlinear regime. For many nonlinear chains, including the Toda~\cite{TodaBook} and granular chains~\cite{Nesterenko1}, by expanding the nonlinear interaction potential and retaining up to cubic nonlinear term we obtain the FPUT system. To assess the universality of the recurrence observed in this work, we considere two different models: the diatomic Toda and granular chains in the weakly nonlinear regime. To investigate the effect of higher-order terms on recurrence, we deal with the full potential instead of an approximated potential for which we retain terms up to certain order. 
For the Toda lattice, interaction potential is given by~\cite{TodaBook}
\begin{equation}
\phi(r)= \frac{a}{b}(\mathrm{e}^{-br}-1)+ar,
\end{equation}
where $a$ and $b$ are constants.

For the granular chains we conider the following potential:
by~\cite{Nesterenko1} 
\begin{eqnarray}\phi(r)=
\begin{cases}
	c(\Delta-r)^{2.5}, &r\leq\Delta \cr 0, &r>\Delta \end{cases},~\quad c = \mathrm{constant},
\label{e:potential}
\end{eqnarray}
where $c$ is constant, and $\Delta$ is the precompression induced by external force.  

\begin{figure}[htbp]
\centerline{
\includegraphics[width=0.49\textwidth]{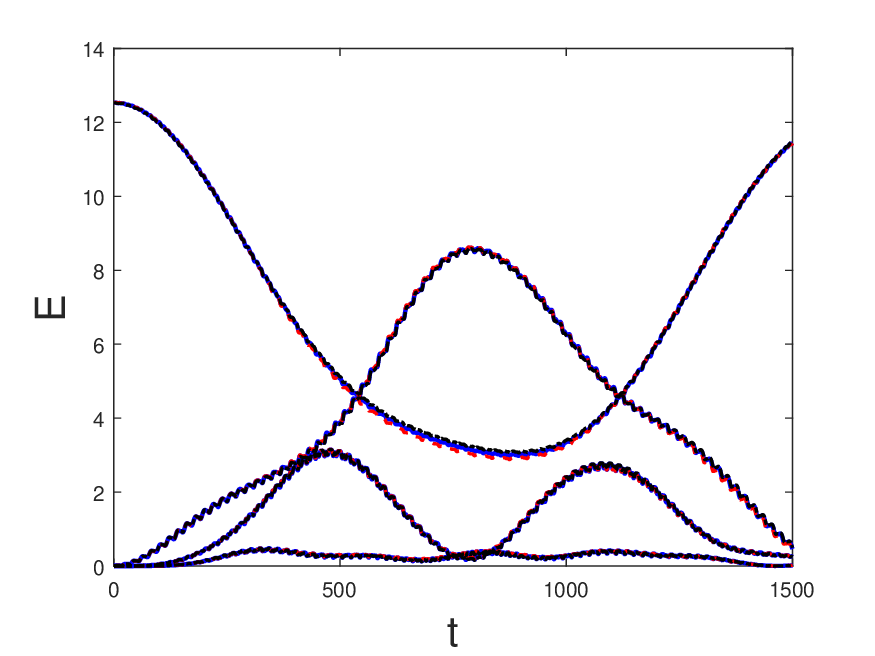}
\includegraphics[width=0.49\textwidth]{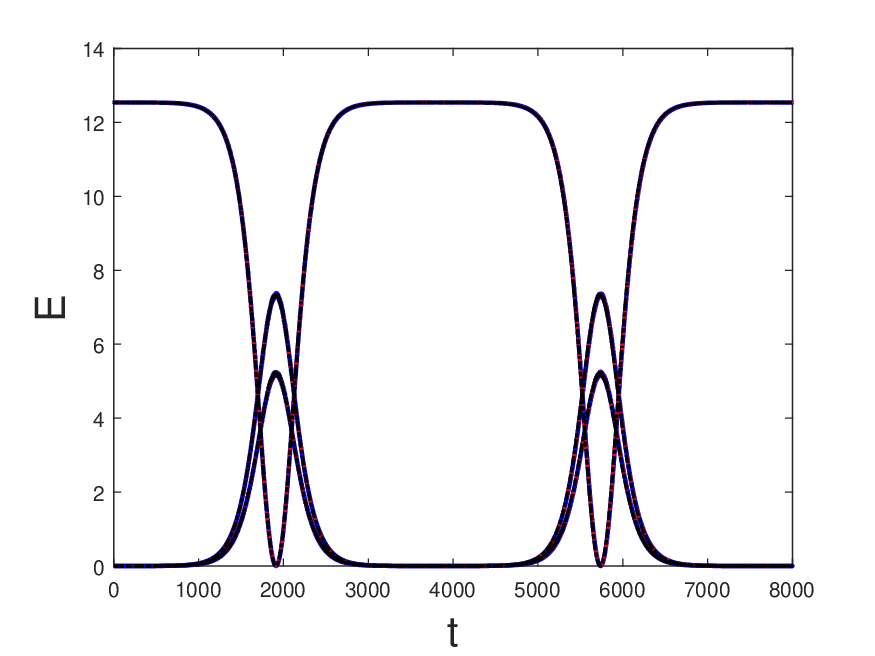}
}
\caption{Recurrence of both types in the FPUT (blue), Toda (red) and granular (black) systems.
}
\label{f:models}
\end{figure}

For comparison purposes, we set parameters in either Toda or granular chains in a way so that the coefficients of both the harmonic and cubic terms in the expansion of the interaction potential are the same as those in  our FPUT model, i.e., with $\kappa=1$ and $\alpha=1$ in~\eqref{cubicpot1}. We then compare the recurrence observed in these models with that found in the classical and diatomic FPUT lattices. Figure~\ref{f:models} shows examples of both types of recurrence. We find that the inclusion of higher-order terms has only a minor effect on the recurrent dynamics, at least over the timescales in our simulations.

\subsection{Effect of fluctuation in mass on recurrence}
As shown in Section~\ref{s:recurrence due to resonance}, the recurrence arising from resonant interactions among Fourier modes requires the mass ratio of the diatomic chain to satisfy a specific resonance condition. In practice, however, the particle masses may not take their nominal values exactly and can exhibit small fluctuations. It is therefore important to investigate whether this resonance-induced recurrence remains observable in the presence of mass disorder.

Assuming the mass of both light and heavy particles are subjected to Gaussian distribution, where the average value is given by $m_{1,2}$ and the variance is given by $\sigma\,m_{1,2} /3$. Then $99.73\%$ of the particle mass are in the range $[m_{1,2}(1-\sigma),\,\,m_{1,2}(1+\sigma)]$.

\begin{figure}[htbp]
\centerline{
\includegraphics[width=0.49\textwidth]{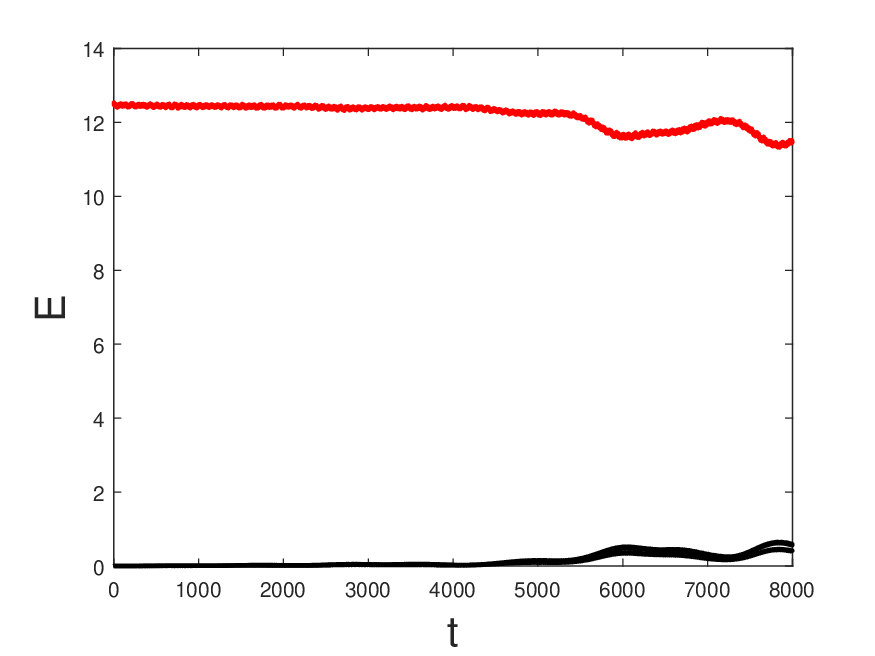}
\includegraphics[width=0.49\textwidth]{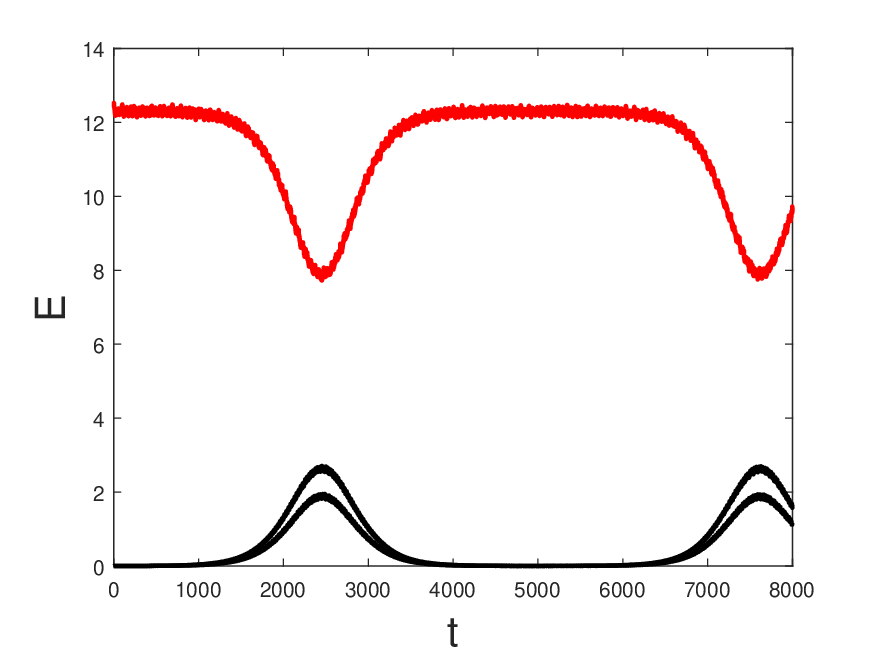}
}
\caption{The energy of Fourier modes as a function of time in diatomic FPUT lattice with variances in mass. Left: no obvious recursive behavior. Right: incomplete recursive behavior.
}
\label{f:fluc}
\end{figure}

We investigate a perturbed diatomic chain, in which the average mass is given by $m_{1,2}=1,\, 1.9831$, and the variance in mass is given by $0.01\,m_{1,2} /3$, $0.05\,m_{1,2} /3$ and $0.1\,m_{1,2}/3$ respectively. Note that the mass of particles is no longer a fixed number, instead it is a random number subjected to normal distribution. Hence we do not have deterministic results, to obtain the distribution of our results we run 100 simulations for each parameter set.

\begin{table}[htbp]
\caption{Distribution of simulation results.}
\label{t:1}
\centering
\begin{tabular}{|c |c | c | c |} \hline
 Variance $\sigma$ & Recurrence &  Incomplete recurrence & No recurrence  \\ \hline
0.01      & 98  & 2      & 0  \\ \hline
0.05      & 81  & 16     & 3 \\ \hline
0.1       & 57  & 27     & 16 \\ \hline

\end{tabular}
\end{table}

Here following~\cite{Porter2023}, we define a recurrence is complete if it satisfies the following two conditions: (i) the energy flows from the exciting Fourier mode to other Fourier modes is more than $2/3$ of the system energy; (ii) after the first cycle more than $2/3$ of the system energy returns to the initially excited Fourier mode. An incomplete recurrence is such that (i) the energy flows from the exciting Fourier mode to other Fourier modes is between $1/3$ and $2/3$ of the system energy; (ii) after the first cycle the portion of the system energy returning to the initially excited Fourier mode is between $1/3$ and $2/3$. Otherwise there is no recurrence. 

For particle numbers $2N=32$, nonlinear strength $\alpha=0.015$, due to variances in mass we do not observe obvious recursive behavior in some of our trials as shown in the left panel of Figure~\ref{f:fluc}. We also observe incomplete recurrence in some trials as shown in the right panel of Figure~\ref{f:fluc}.
Nevertheless we still observe complete recurrence in a large amount of trials. Distribution of our results is reported in Table~\ref{t:1}.

\section{Reduction to the three-wave integrable system }\label{Sec:threewave}

It is worth noting that, in the thermodynamic limit $N \to \infty$, and under the assumption that the Fourier spectrum remains narrow-banded in the vicinity of each resonant wavenumber, Eq.~\eqref{e:motion in Fourier} can be recast as a system of three integrable partial differential equations, constituting a universal model for the dynamics of three-wave systems.
In this limit, $k$ can be treated as a continuous variable,  consequently $a^{\pm}=a^{\pm}(k,t)$ becomes a function of two variables. Assuming that $k_A,k_B,k_C$  are three resonant wavenumbers, we can write  $a^{\pm}(k,t)$ as
\begin{equation}
\begin{split}
   & a^+ (k,t)=A(k-k_A,t) e^{(-i \omega^+(k_A)t)},\\
   & a^- (k,t)=B(k-k_B,t) e^{(-i \omega^-(k_B)t)}+C(k-k_C,t) e^{(-i \omega^-(k_C)t)}.
\end{split}
\label{e:expand a}
\end{equation}
Substituting~\eqref{e:expand a} into~\eqref{eq:3wo} 
in the continuous limit, where summations become integrals, retaining only the resonant contributions, we obtain the following evolution equations for $A$, $B$ and $C$:
\begin{equation}
\begin{split}
  &i\frac{\partial A}{\partial t}= \frac{d \omega^+}{dk}\bigg|_{k=k_A}(k-k_A) A
  +2\int_{-\infty}^{+\infty}\bar{V}_{A,B,C}^{(2)} B(k_1-k_B,t)C(k-k_1-k_C,t)\, dk_1,\\
  &i\frac{\partial B}{\partial t}=\frac{d \omega^-}{dk}\bigg|_{k=k_B} (k-k_B)B
  +2\int_{-\infty}^{+\infty}\bar{V}_{A,B,C}^{(2)*}A(k_1-k_A,t)C^*(k_1-k-k_C,t)\, dk_1,\\
  &i\frac{\partial C}{\partial t}=\frac{d \omega^-}{dk}\bigg|_{k=k_C} (k-k_C)C
  +2\int_{-\infty}^{+\infty}\bar{V}_{A,B,C}^{(2)*}A(k_1-k_A,t)B^*(k_1-k-k_B,t)\, dk_1.
\end{split}
\label{eq:ABC}
\end{equation}

We introduce $\widetilde{A}$, $\widetilde{B}$ and $\widetilde{C}$ as the inverse Fourier transform of $A$, $B$ and $C$ respectively. Explicitly they are defined as
\begin{equation}
\begin{split}
& \widetilde{A}=\int_{-\infty}^{+\infty} A\mathrm{e}^{i(k-k_A)x}dk,\\
& \widetilde{B}=\int_{-\infty}^{+\infty} B\mathrm{e}^{i(k-k_B)x}dk,\\& 
\widetilde{C}=\int_{-\infty}^{+\infty} C\mathrm{e}^{i(k-k_C)x}dk.
\end{split}
\end{equation}
Applying the inverse Fourier transform to~\eqref{eq:ABC}, taking into account that $\,\bar V_{A,B,C}^{(2)}\,$ is purely imaginary and recalling the convolution theorem, we obtain

\begin{equation}
\begin{split}
  &\frac{\partial \widetilde{A}}{\partial t}- \frac{d \omega^+}{dk}\bigg|_{k=k_A}\frac{\partial \widetilde{A}}{\partial x}
  =-2i\bar{V}_{A,B,C}^{(2)}\widetilde{B}\widetilde{C},\\
  &\frac{\partial \widetilde{B}}{\partial t}-\frac{d \omega^-}{dk}\bigg|_{k=k_B} \frac{\partial \widetilde{B}}{\partial x}
  =2i\bar{V}_{A,B,C}^{(2)}\widetilde{A}\widetilde{C}^*,\\
  &\frac{\partial \widetilde{C}}{\partial t}-\frac{d \omega^-}{dk}\bigg|_{k=k_C} \frac{\partial \widetilde{C}}{\partial x}
  =2i\bar{V}_{A,B,C}^{(2)}\widetilde{A}\widetilde{B}^*.
\end{split}
\label{eq:ABC1}
\end{equation}

\textcolor{black}{
Eq.~\eqref{eq:ABC1} is an integrable system, representing the universal model for the evolution of an isolated three–wave resonant triad (the classical \textit{3WRI system}), i.e., a triad dynamically decoupled from the surrounding resonance network \cite{zakharov1973resonant,zakharov1975theory, kaup1978applications}.}

\section{Conclusion}
\label{s:conclusion}
In this work, we have investigated recurrence phenomena in the diatomic $\alpha$--FPUT lattice and identified two distinct mechanisms leading to recurrent energy exchange. The first corresponds to the classical FPUT recurrence, which was originally observed in monatomic anharmonic chains and is shown here to persist in the diatomic setting. Consistent with previous studies, we find that the corresponding recurrence period scales approximately as $1/\sqrt{\alpha}$, where $\alpha$ denotes the nonlinear strength.
The second type of recurrence is specific to the diatomic lattice and originates from exact three-wave resonant interactions involving two acoustic modes and one optical mode. We have shown that such resonances occur only for specific values of the mass ratio and have derived a criterion that determines whether a given triplet of Fourier modes can satisfy the resonance conditions. Numerical simulations demonstrate that these resonant interactions lead to a robust and long-lived recurrence in which the energy is exchanged almost exclusively within a resonant triad.
Motivated by this observation, we derived a reduced three-mode model in Fourier space and showed that it reproduces the main dynamical features of the full lattice. In the thermodynamic limit, this model reduces to the classical integrable three-wave resonant interaction system, providing a natural theoretical framework for understanding the observed recurrence.

A notable difference between the two recurrence mechanisms emerges in the weakly nonlinear limit. As $\alpha$ decreases, the period of both recurrences increases. However, while the classical FPUT recurrence becomes progressively weaker and eventually difficult to observe, the resonance-induced recurrence remains essentially unaffected, apart from a rescaling of the characteristic time scale. This distinction highlights the fundamentally different origin of the two phenomena.

We have also examined the robustness of the recurrence by considering two different nonlinear interaction potentials, namely the Toda and granular chains, as well as random fluctuations in the particle masses. Our study has illustrated that including higher-order terms has a negligibly small effect on recursive behavior. Then, we have considered  the effect of fluctuations in the particle masses.
Our results suggest that the recurrence behavior is still observable for systems with relatively small mass fluctuations despite its specific requirement on mass ratio.

The present work raises several interesting questions for future investigation. In particular, it would be valuable to understand how the resonance-induced recurrence evolves in larger systems, where resonant triads become embedded in extended resonance networks, and to determine under which conditions the phenomenon may be observed experimentally. More generally, the interplay between recurrence, resonant energy transfer, and thermalization in perturbed lattices remains an open and fascinating problem.
\\

{\bf Acknowledgments}.  G. D. was funded by National Natural Science Foundation of China
(No. 12501329) and Guangzhou GI Grant SL2024A04J02041. G. L. was supported by National Natural Science Foundation of China (12371162 and 12001127),
 Guangdong Basic and Applied Basic Research Foundation (2024A1515010867), and Science and Technology Program of Guangzhou (202201020546 and 2025A04J5160). 
 M.O. was supported by INFN through the MMNLP and FIELDTUR projects and  by the Simons Foundation (USA) under Awards No. 652354 and  No. 651471 (Wave Turbulence).

\bibliography{reference.bib}

@article{Bambusi1993,
  title={Exponential stability of states close to resonance in infinite-dimensional {H}amiltonian systems},
  author={D. Bambusi and A. Giorgilli},
  journal={J. Stat. Phys.},
  year={1993},
  volume={71},
  pages={569-606}
}

@article{pezzi2025multi,
  title={Multi-wave resonances in the diatomic $\alpha$-FPUT system},
  author={Pezzi, A and Deng, G and Lvov, Y and Lorenzo, M and Onorato, M},
  journal={Chaos, Solitons \& Fractals},
  volume={192},
  pages={116005},
  year={2025},
  publisher={Elsevier}
}

@article{Benettin2013,
title={The {F}ermi-{P}asta-{U}lam problem and its underlying integrable dynamics},
author={G. Benettin and H. Christodoulidi and A. Ponno},
journal={J. Stat. Phys.},
volume={152},
pages={195–212},
year={2013}
}

@article{ZhangZhao,
  title = {Diffusion of heat, energy, momentum, and mass in one-dimensional systems},
  author = {S. Chen and Y. Zhang and J. Wang and H. Zhao},
  journal = {Phys. Rev. E},
  volume = {87},
  issue = {3},
  pages = {032153},
  numpages = {7},
  year = {2013}
}

@article{fermi1955,
  title={{L}os {A}lamos {R}eport {LA}-1940},
  author={E. Fermi and J. Pasta and S. Ulam},
  journal={E. Fermi, Collected Papers},
  volume={2},
  pages={977--988},
  year={1955}
}

@article{FrieseckePego,
  title={Solitary 
waves on {FPU} 
lattices: {I}. {Q}ualitative properties, renormalization and continuum limit},
  author={G. Friesecke and R. L. Pego},
  journal={Nonlinearity},
  volume={12},
  number={6},
  pages={1601--1627},
  year={1999}
}

@article{Fu2019,
year = {2019},
volume = {21},
number = {4},
pages = {043009},
author = {W. Fu and Y. Zhang and H. Zhao},
title = {Universal law of thermalization for one-dimensional perturbed {T}oda lattices},
journal = {New J. Phys.}
}

@article{FrieseckeWattis,
  title={Existence theorem for solitary waves on lattices},
  author={G. Friesecke and J. A. D. Wattis},
  journal={Commun. Math. Phys.},
  volume={161},
  number={2},
  pages={391--418},
  year={1994}
}

@article{Galgani1992,
  title={On the problem of energy equipartition for large systems of the {F}ermi-{P}asta-{U}lam type: analytical and numerical estimates},
  author={L. Galgani and A. Giorgilli and A. Martinoli and S. Vanzini},
  journal={Physica D},
  year={1992},
  volume={59},
  pages={334-348}
}

@book{lepri2016thermal,
  title={Thermal transport in low dimensions: from statistical physics to nanoscale heat transfer},
  author={S. Lepri},
  volume={921},
  year={2016},
  publisher={Springer}
}

@article{Maiocchi2019,
    author = {A. M. Maiocchi},
    title = "{Freezing of the optical-branch energy in a diatomic {FPU} chain}",
    journal = {Commun. Math. Phys.},
    volume = {372},
    pages = {91–117},
    year = {2019}
}

@article{Porter2023,
    author = {Z. Li and M. A. Porter and B. Choubey},
    title = "{Recurrence recovery in heterogeneous {F}ermi–{P}asta–{U}lam–{T}singou systems}",
    journal = {Chaos: {A}n {I}nterdisciplinary {J}ournal of {N}onlinear {S}cience},
    volume = {33},
    number = {9},
    pages = {093108},
    year = {2023}
}

@article{lvov2018,
  title={Double scaling in the relaxation time in the $\beta$-{F}ermi-{P}asta-{U}lam-{T}singou model},
  author={Y. V. Lvov and M. Onorato},
  journal={Phys. Rev. Lett.},
  volume={120},
  number={14},
  pages={144301},
  year={2018}}

@book{nazarenko2011,
  title={Wave turbulence},
  author={S. Nazarenko},
  volume={825},
  year={2011},
  publisher={Springer Science \& Business Media}
}

@article{onorato2015,
  title={Route to thermalization in the $\alpha$-{F}ermi-{P}asta-{U}lam system},
  author={M. Onorato and L. Vozella and D. Proment and Y. V. Lvov},
  journal={Proceedings of the National Academy of Sciences},
  volume={112},
  number={14},
  pages={4208--4213},
  year={2015}
}

@article{Campbell2019,
    author = {S. D. Pace and K. A. Reiss and D. K. Campbell},
    title = "{The $\beta$ {F}ermi-{P}asta-{U}lam-{T}singou recurrence problem}",
    journal = {Chaos: {A}n {I}nterdisciplinary {J}ournal of {N}onlinear {S}cience},
    volume = {29},
    number = {11},
    pages = {113107},
    year = {2019}
}

@book{Nesterenko1,
  title={Dynamics of {H}eterogeneous {M}aterials},
  author={V. F. Nesterenko},
  year={2001},
  publisher={Springer-Verlag},
  address = {Heidelberg, Germany}
}

@article{Pezzi2021,
    author = {A. Pezzi and G. Deng and Y. Lvov and M. Lorenzo and M. Onorato},
    title = "{Three-wave resonant interactions in the diatomic chain with cubic anharmonic
    potential: theory and simulations}",
    journal = {arxiv: 2103.08336},
    year = {2021}
}

@article{pistone2018,
  title={Thermalization in the discrete nonlinear {K}lein-{G}ordon chain in the wave-turbulence framework},
  author={L. Pistone and M. Onorato and S. Chibbaro},
  journal={Europhys. Lett.},
  volume={121},
  number={4},
  pages={44003},
  year={2018}
}

@book{TodaBook,
  title={Theory of {N}onlinear {L}attices},
  author={M. Toda},
  year={1989},
  publisher={Springer-Verlag}
}

@article{ZabuskyKruskal1965,
  title = {Interaction of "Solitons" in a Collisionless Plasma and the Recurrence of Initial States},
  author = {N. J. Zabusky and M. D. Kruskal},
  journal = {Phys. Rev. Lett.},
  volume = {15},
  issue = {6},
  pages = {240--243},
  numpages = {0},
  year = {1965}
}

@book{zakharov2012,
  title={Kolmogorov spectra of turbulence I: Wave turbulence},
  author={V. E. Zakharov and V. S. L'vov  and G. Falkovich},
  year={2012},
  publisher={Springer Science \& Business Media}
}

@article{bustamante2011resonance,
  title={Resonance clustering in wave turbulent regimes: Integrable dynamics},
  author={Bustamante, Miguel D and Kartashova, Elena},
  journal={Communications in Computational Physics},
  volume={10},
  number={5},
  pages={1211--1240},
  year={2011},
  publisher={Cambridge University Press}
}

@techreport{zakharov1973resonant,
  title={Resonant interaction of wave packets in nonlinear media},
  author={Zakharov, VE and Manakov, SV},
  year={1973},
  institution={Inst. of Nuclear Physics, Novosibirsk, USSR}
}

@article{zakharov1975theory,
  title={The theory of resonance interaction of wave packets in nonlinear media},
  author={Zakharov, VE and Manakov, SV},
  journal={Zh. Eksp. Teor. Fiz},
  volume={69},
  number={5},
  pages={1654--1673},
  year={1975}
}

@article{kaup1978applications,
  title={Applications of the inverse scattering transform II: the three-wave resonant interaction},
  author={Kaup, DJ},
  journal={The Rocky Mountain Journal of Mathematics},
  pages={283--308},
  year={1978},
  publisher={JSTOR}
}
\bibliographystyle{unsrt}

\newpage
\appendix
\section{Calculation of the coefficients}
\label{AppendixA}
The coefficients of the transformed Hamiltonian \eqref{FullHamiltonian} are
\begin{subequations}
\label{e:W}
\label{coefficients}
\begin{align}
W_{1,2,3}^{(1)} &= [A_{1,2,3}^{+++}+A_{2,1,3}^{+++}+A_{3,2,1}^{+++}]/3,
\label{coefficients_a} \\
W_{1,2,3}^{(2)} &=  [A_{1,2,3}^{---}+A_{2,1,3}^{---}+A_{3,2,1}^{---}]/3,
\label{coefficients_b} \\
W_{1,2,3}^{(3)} &=A_{1,2,3}^{+--} + A_{2,1,3}^{-+-} + A_{3,1,2}^{-+-},
\label{coefficients_c}\\
W_{1,2,3}^{(4)} &= A_{1,2,3}^{-++} +A_{2,1,3}^{+-+} + A_{3,1,2}^{+-+},
\label{coefficients_d}
\end{align}
\end{subequations}
with
\begin{equation}
A_{1,2,3}^{s_1,s_2,s_3}=(\beta_{s_1,k_1} + (-1)^l \beta_{s_2,k_2}\beta_{s_3,k_3} )\sin(aq_{k_1})
\end{equation}
and $l=(k_2+k_3+k_1)/N$.

The coefficients in~\eqref{nlsystemdiagonal} are 
    	\begin{subequations}
		\label{3.90}
		\begin{align}
			V_{1,2,3}^{(1)}&=-3 W_{-1,2,3}^{(1)}, \qquad
			V_{1,2,3}^{(2)}=- W_{-1,2,3}^{(3)}, \qquad
			V_{1,2,3}^{(3)}= -2W_{3,-1,2}^{(4)}, \\
			T_{1,2,3}^{(1)}&=-W_{-1,2,3}^{(4)}, \qquad
			T_{1,2,3}^{(2)}=-3W_{-1,2,3}^{(2)}, \qquad
			T_{1,2,3}^{(3)}=-2W_{2,-1,-3}^{(3)},
		\end{align} 
	\end{subequations}
where $W_{k_1,k_2,k_3}^{(i)}$ are given in~\eqref{e:W}.

The coefficients in~\eqref{eq:3wo} are 
    \begin{subequations}
		\begin{align}
			&\bar{V}_{1,2,3}^{(1)}=\gamma_{1,2,3}^{+ + +}V_{1,2,3}^{(1)}, \;\;\;
			\bar{V}_{1,2,3}^{(2)} =\gamma_{1,2,3}^{+ --}V_{1,2,3}^{(2)}, \;\;\;
			\bar{V}_{1,2,3}^{(3)} =\gamma_{1,2,3}^{+ +-}V_{1,2,3}^{(3)}, \;\;\; \\
			&\bar{T}_{1,2,3}^{(1)}=\gamma_{1,2,3}^{- + +}T_{1,2,3}^{(1)}, \;\;\;
			\bar{T}_{1,2,3}^{(2)} =\gamma_{1,2,3}^{- --}T_{1,2,3}^{(2)}, \;\;\;
			\bar{T}_{1,2,3}^{(3)} =\gamma_{1,2,3}^{- +-}T_{1,2,3}^{(3)}, \;\;\;
		\end{align}
	\end{subequations}
with
	\begin{equation}
		\gamma_{1,2,3}^{s_1 s_2 s_3}=
        -\frac{i\alpha}{N\sqrt{2m_{12,k_1}^{s_1}m_{12,k_2}^{s_2}m_{12,k_3}^{s_3}\omega_{k_1}^{s_1}\omega_{k_2}^{s_2}\omega_{k_3}^{s_3}}}.
	\end{equation}

\end{document}